\documentclass[a4paper,11pt]{article}
\usepackage{jcappub} 
\usepackage{lineno}

\title{\boldmath Increasing the Precision of Surrogate Models for Weak Lensing Mass Maps with Flow Matching}

\author[a]{Guangjian Li}
\author[b]{Tomasz Kacprzak}

\affiliation[a]{\small Institute of Astrophysics, University of Zurich,\\
Winterthurerstrasse 190, 8057 Zurich, Switzerland}

\affiliation[b]{\small Institute for Data Science, University of Applied Sciences and Arts Northwestern Switzerland, \\
Bahnhofstrasse 6, 5210 Windisch, Switzerland}

\emailAdd{guangjian.li@uzh.ch, tomasz.kacprzak@fhnw.ch}

\abstract{Weak gravitational lensing maps compactly encode the evolution of cosmic large-scale structure and have become a key tool for cosmological analyses in astronomical surveys. 
Performing inference directly on the map-level allows for flexibility in the choice of statistics, which can lead to increased constraining power from the same data. 
Conventional methods rely solely on N-body simulations and are computationally expensive. 
Generative machine learning emulators, fitted to relatively small N-body training sets, can accelerate map-level theory prediction. 
However, existing map-level surrogates, such as those based on generative adversarial networks (GAN), have not yet reached the quality needed for their practical application in survey analyses: they match summaries with limited precision, produce over-smooth maps, have difficulties in capturing the distribution of generated map sets correctly, and can be difficult to train.
Continuous normalizing flows trained with flow matching have recently emerged as a powerful method for generative artificial intelligence.
We present a residual label-conditional flow matching (RLCFM) generative network that conditions explicitly on the matter density $\Omega_m$ and clustering amplitude $\sigma_8$ for a fixed source redshift distribution $n(z)$. 
The model learns a continuous probability flow in a residual space from label-specific noise distributions to convergence maps. 
We evaluate it using a comprehensive set of validation metrics, including pixel and peak statistics, the power spectrum, bispectrum, correlation matrices of the power spectra, and others. 
Our RLCFM method greatly improves on the previous GAN benchmark, increasing the typical fidelity of generated maps from previous levels of $<10\%$ ($<20\%$) to $<1\%$ ($<5\%$) for basic (higher-order) statistics.
Importantly, the agreement at the level of map distributions is very good: the distribution of maps generated with our surrogate from random noise matches well with the distribution of maps generated with N-body code from random initial conditions.
This work brings us closer to a practical mass map emulator that captures the cosmological signal while supporting multiple forms of data analysis.

\keywords{weak gravitational lensing, mass maps, convergence fields, conditional generative models, normalizing flows, flow matching, cosmological simulations}

} 

\usepackage{amsthm}
\usepackage{thmtools}
\usepackage{thm-restate}

\usepackage{booktabs}

\begin{document}
\maketitle
\flushbottom

\section{Introduction}

Weak gravitational lensing is the small distortion of distant galaxy shapes caused by the matter distribution along the line-of-sight. It has become a key probe in modern cosmology \cite{bartelmann2001weak, kilbinger2015cosmology}. Because lensing is directly sensitive to the projected total matter distribution and does not rely on galaxy bias, cosmic shear measurements provide a powerful probe of the growth of structure and the expansion history of the Universe. Current and future wide-field weak-lensing surveys, such as KiDS \citep{de2013kilo}, DES \citep{dark2016dark}, HSC \citep{aihara2018first}, Euclid \citep{laureijs2011euclid}, and the Rubin Observatory LSST \citep{ivezic2019lsst}, aim to measure key cosmological parameters, such as the matter density $\Omega_{\mathrm{m}}$, the amplitude of matter fluctuations $\sigma_8$, and the properties of dark energy, with percent-level precision.

In weak-lensing analyses, projected mass maps (e.g., convergence fields) provide a direct view of the matter distribution and enable the extraction of non-Gaussian information beyond two-point statistics. Accurate forward modeling of such maps typically relies on large-volume, high-resolution N-body simulations coupled with ray tracing \cite{jain2000ray, takahashi2017full}. These simulations capture non-linear evolution, scale coupling, and non-Gaussian features that strongly affect higher-order observables such as bispectrum and Minkowski functionals. However, each simulation is computationally expensive, and exploring multi-dimensional cosmological parameter space requires many realizations \citep{petri2016sample, harnois2019cosmic}. 

Deep learning has recently emerged as a powerful framework for working with weak-lensing data. 
Convolutional neural networks (CNNs) have been used to infer cosmological parameters directly from mass maps, often outperforming traditional two-point statistics \citep{ribli2019weak,fluri2022full,sharma2024comparative}. 
For map generation, generative adversarial networks (GANs) have been developed to emulate weak-lensing convergence maps\citep{mustafa2019cosmogan,perraudin2021emulation,yiu2022tomographic}.
Diffusion and score-based methods have also been applied to weak-lensing mass map denoising and reconstruction \citep{remy2023probabilistic,aoyama2025denoising,boruah2025diffusion}.

Flow-based models have recently emerged as a powerful engine for generative artificial intelligence (AI).
They learn continuous transports between probability distributions, enabling efficient sampling and conditional generation\citep{dinh2016density,kingma2018glow,papamakarios2021normalizing}. 
In particular, flow matching \citep[FM, ][]{lipman2022flow,liu2022flow} trains continuous normalizing flows \citep[CNF, ][]{chen2018neural} by directly regressing the time-dependent vector field that transports a prior to the data distribution. 
This formulation turns training into a supervised regression problem and reduces the need to repeatedly solve ODE trajectories during optimization. 
CNF-FM achieves competitive performance relative to diffusion models while enabling efficient sampling \citep{lipman2022flow,tong2023improving}. 
Recent studies have explored flow matching for field-level cosmological applications. Diao et al. \citep{diao2025detecting} investigated continuous-time flow models for identifying modeling biases in weak-lensing maps, while Kannan et al. \citep{kannan2025cosmoflow} developed CosmoFlow, which explores scale-aware representation learning for cosmological data using flow matching.
Of particular relevance to our work, Zeghal et al. \citep{zeghal2025bridging} proposed conditional optimal transport flow matching (COT-FM) to bridge low- and high-fidelity weak-lensing simulations. Their approach uses maps generated by second-order Lagrangian perturbation theory (2LPT) as low-fidelity inputs and high-fidelity N-body Particle-Mesh (PM) maps as targets, with minibatch optimal transport used to construct sample couplings between the two datasets. They further demonstrated the utility of the resulting emulator in a full-field cosmological inference pipeline, obtaining posterior constraints in close agreement with those derived using the high-fidelity simulations.

Despite recent progress, accurately reproducing the full conditional distribution of weak-lensing maps across cosmological parameter space remains challenging. Existing GAN-based emulators provide a useful benchmark for weak-lensing map generation \citep{perraudin2021emulation,yiu2022tomographic}.
They show agreement on the central values of summary statistics at the 5\%-10\% level, when averaged over initial conditions.
However, their performance can vary across cosmological parameter space, with larger discrepancies in non-Gaussian statistics and near the edges of the training grid. Moreover, reproducing the distributions of summary statistics across random initial conditions remains challenging, suggesting that agreement in central values does not necessarily imply accurate recovery of the full statistical variability of the maps.
This showed that GANs were able to fit the ``mode'' of the map properties, but not their full probabilistic distribution.
These works also showed that GAN-based mass map emulators can be computationally demanding to train and sensitive to training instabilities, which may limit their practical scalability \citep{tamosiunas2021investigating}. 
More broadly, conditional generation across continuous labels is inherently challenging, since the target distribution $q(x|y)$ can vary substantially with $y$, leading to heterogeneous learning difficulty across the label space. 
In the context of flow-based generative modeling, recent studies on prior design suggest that adopting a more informative, data-driven source distribution that is closer to the corresponding conditional targets can simplify the learned transport and improve training stability and sampling efficiency across diverse domains \citep{kollovieh2024flow,issachar2025designing,wu2025flowdesign}. 
Motivated by these observations, we develop a conditional flow-matching emulator and evaluate it through extensive experiments, with a particular focus on interpolation performance at held-out test cosmologies.

In this work, we present a map-level emulator for weak-lensing convergence that is based on conditional flow matching and trained on suites of N-body simulations. 
We start by designing a simple, data-driven Gaussian prior for the convergence field, then learn a label-conditional continuous flow in a residual space that transports this prior into realistic mass maps at given cosmological parameters $\Omega_m$ and $\sigma_8$. 
Unlike the simulator-bridging approach of Zeghal et al. \citep{zeghal2025bridging} that learns a conditional transport between low- and high-fidelity simulation distributions, our approach uses predicted conditional mean and scale as a reference for residual-space generation. It therefore requires neither an additional low-fidelity simulation suite nor paired low- and high-fidelity map realizations during training.
After training, the model can rapidly generate large ensembles of mass maps and smoothly interpolate to unseen cosmological parameters, while being significantly cheaper than N-body simulations and requiring substantially shorter training times. 
We show that the RLCFM emulator reproduces key statistics of the simulation maps, such as pixel distributions, power spectra, and non-Gaussian observables such as peak counts, bispectrum, and Minkowski functionals, across cosmological parameter space.
Our results demonstrate improved statistical accuracy and interpolation performance compared with the GAN baseline considered in this work.
Performance remains stable at both training and test cosmologies, enhancing the practical usability of the model for forward modeling applications. 
These results demonstrate that conditional flow matching is a useful tool for precise weak-lensing forward modeling and a step toward fully generative, simulation-based cosmological inference pipelines.

This paper is structured as follows. Section~\ref{sec:rlcfm} introduces the residual label-conditional flow matching model. Section~\ref{sec:dataset} describes the dataset used for training. Section~\ref{sec:metrics} outlines the validation metrics. 
Section~\ref{sec:results} presents the generated maps and compares them to the N-body simulations. 
Conclusions and future directions are given in Section~\ref{sec:conclusion}. Appendix~\ref{app:network} details the neural network architectures,
Appendix~\ref{app:ablation} shows a comparison with the existing GAN model results and ablation experiments. 
Appendix~\ref{app:proof} provides proofs of the main theorems and the reparameterization of the Gaussian probability path.
\section{Conditional flow matching}\label{sec:rlcfm}

In this section, we give an introduction to the standard continuous normalizing flow and flow matching models, as well as our proposed extension to the residual label-conditional model.

\subsection{Continuous normalizing flows}
Generative models aim to learn the underlying data distribution and enable drawing new samples from it. 
Continuous Normalizing Flows \cite[CNF, ][]{chen2018neural} realize this via a flow-based approach that learns a neural velocity field defining an ordinary differential equation (ODE), thereby transporting a base distribution into the target distribution. 
Let $R^n$ denote the data space with data points $x=(x^{(1)},\ldots,x^{(n)})\in R^n$, we can define a time-dependent probability path $p:[0,1]\times R^n \to R_{>0}$ satisfying $\int p_t(x) dx=1$ for all $t\in[0,1]$, and a time-dependent probability vector field $u:[0,1]\times R^n \to R^n$. 
A valid probability transport induced by the velocity field must satisfy the continuity equation
\begin{equation}
\partial_t p_t(x)+\nabla_x\cdot(p_t(x)u_t(x))=0,
\label{eq:continuity}
\end{equation}
which describes the evolution of the probability density along the transport path. 
The time-dependent flow $\phi:[0,1]\times R^n \to R^n$ is then induced by $u_t(x)$ via an ODE:
\begin{equation}
\frac{d}{dt}\phi_t(x)=u_t\big(\phi_t(x)\big),\quad \phi_0(x)=x.
\label{eq:OED}
\end{equation}
The vector field can be parameterized as a neural network $v_{\theta}(t,x)$ with learnable parameters $\theta$ \cite{chen2018neural}. 
Eq.~\eqref{eq:OED} then determines the associated flow $\phi_t(x)$, thereby yielding a CNF. 
Given a prior distribution $p_0$ at $t{=}0$, the forward dynamics induce a time-dependent distribution $p_t$ via the push-forward relation \cite{lipman2022flow}:
\begin{equation}
p_t(x)=[\phi_t]_*\, p_0(x)
= p_0\!\big(\phi_t^{-1}(x)\big)\,
\det\!\left[\frac{\partial \phi_t^{-1}}{\partial x}(x)\right].
\label{eq:push}
\end{equation}
In particular, if $p_1$ at $t=1$ is the desired target distribution, the flow transports $p_0$ to $p_1$.
The goal in CNF is to train the neural network $v_{\theta}(t,x)$ such that the ODE induced flow $\phi_t(x)$ (Eq.~\eqref{eq:OED}) maps the base distribution $p_0$ onto the target $p_1$ (Eq.~\eqref{eq:push}). 
One key challenge of classic CNFs is high computational cost: the neural network $v_t(x;\theta)$ learns the target velocity field indirectly by maximizing likelihood along the continuous flow, and both training and sampling require numerous ODE solver evaluations, which may increase during training and limit scalability.

\subsection{Flow matching}
Lipman et al.\ (2022) proposed Flow Matching~\cite{lipman2022flow}
 (FM), an efficient method for regressing velocity fields directly. 
 FM is a simulation-free approach for training CNFs by regressing time-dependent vector fields explicitly without frequent ODE evaluations, thereby accelerating convergence and reducing the overall training time. 
Given a predefined probability path $p_t$ connecting a source distribution 
$p_0$ to a target distribution $p_1$, FM learns a neural velocity field 
$v_\theta(t,x)$ that approximates the corresponding target velocity field 
$u_t(x)$. 
The FM objective is formulated as
\begin{equation}
\mathcal L_{\mathrm{FM}}(\theta) = \mathbb{E}_{t\sim \mathcal U[0,1],\  p_t(x)} \,\big\| v_{\theta}(t,x) - u_t(x) \big\|^2.
\end{equation}
In practice, the path $p_t$ and its inducing velocity $u_t$ are typically unknown, and $u_t$ rarely admits a closed form. 
Consequently, the FM objective cannot be applied directly.
In general, only data samples $x_1$ from some unknown distribution $q(x_1)$ are observed, we therefore introduce a conditional probability path $p_t(x|x_1)$ with boundary conditions: 
$p_0(x|x_1)=p_0(x)$ at $t{=}0$ and a distribution concentrated near $x{=}x_1$ at $t{=}1$, e.g., $p_1(x|x_1) = \mathcal{N}(x|x_1,\sigma^2I)$  with small $\sigma$. 
Let $u_t(x|x_1)$ denote the conditional velocity field that generates $p_t(x|x_1)$ via the continuity equation. 
The Conditional Flow Matching (CFM) objective then trains $v_{\theta}(t, x)$ to match this conditional target by minimizing
\begin{equation}
\mathcal{L}_{\mathrm{CFM}}(\theta)
= \mathbb{E}_{t \sim \mathcal U[0,1], \ q(x_1) ,\  p_t(x|x_1)}
\big\| v_{\theta}(t, x) - u_t(x|x_1) \big\|^2,
\label{eq:CFM}
\end{equation}
up to a constant independent of $\theta$, the CFM and FM losses coincide; hence
$\nabla_{\theta}\mathcal{L}_{\mathrm{FM}}(\theta)=\nabla_{\theta}\mathcal{L}_{\mathrm{CFM}}(\theta)$~\cite{lipman2022flow}.
This equivalence allows us to train the network using samples from the conditional velocity field $u_t(x|x_1)$, while the corresponding population objective has the same gradients as the marginal FM objective for $u_t(x)$, without requiring a closed form expression for $u_t(x)$. 
Moreover, the CFM objective supports any conditional probability path that satisfies the boundary conditions, along with compatible conditional velocity fields, thereby enabling flexible path and target design.

\subsection{Residual label-conditional flow matching}\label{2.3}
In many scientific applications, downstream analysis requires generating label-specific maps. 
For continuous labels in particular, we desire a model that interpolates smoothly across the label space, an ability that is crucial when expensive simulations make dense sampling infeasible. 
A practical FM strategy is to inject label information \(y\) directly into
the neural velocity field \(v_{\theta}(t,x,y)\). If the target distribution is
label-dependent, i.e. \(q(x|y)\), a straightforward label-conditioned
extension of CFM is
\begin{equation}
\mathcal{L}_\mathrm{CFM}^{y}(\theta)
=
\mathbb{E}_{t \sim \mathcal{U}[0,1],\ q(x_1, y),\ p_t(x|x_1)}
\big\| v_{\theta}(t,x,y) - u_t(x|x_1) \big\|^2 
\label{eq:CFM_y}.
\end{equation}
Here, the superscript \(y\) indicates that the neural velocity field is
conditioned on the label \(y\), while the conditional probability path
\(p_t(x|x_1)\) and the target velocity field \(u_t(x| x_1)\) remain
shared across labels. Conceptually, this corresponds to a shared-prior
conditional model: the label \(y\) modulates the learned velocity field, but
all labels are still transported from the same source distribution toward
their corresponding target distributions \(q(x|y)\).

However, in many scientific settings, especially in cosmology and astrophysics, the statistics of the physical field vary substantially with the label (e.g., mean fields and covariances). 
Starting all labels from a single shared source distribution can therefore induce different transport distances across $y$, which often leads to unstable training and non-uniform performance across the parameter space. 
A natural approach is to employ a label-specific source prior so that source samples are, on average, closer to their corresponding targets $q(x|y)$; see Fig.~\ref{label_prior}. 
This design reduces the effective transport length and variability across labels, simplifying the learned transport and improving optimization stability and sampling efficiency across diverse conditional generation tasks \citep{kollovieh2024flow,issachar2025designing}.

Nevertheless, in our interpolation-focused evaluation on test cosmological parameter points, we find that a label-specific prior alone does not improve generalization; instead, it consistently degrades performance across the full set of validation metrics, indicating reduced robustness at interpolation points. 
We provide further discussion and an explanation of this behavior in Appendix~\ref{app:ablation}. 
This motivates an additional representation-level reparameterization: we learn the conditional dynamics in a residual space obtained by per-label mean subtraction, which removes trivial label-dependent mean shifts and yields more stable interpolation across cosmological parameters. 
In Appendix~\ref{app:ablation}, we conduct an ablation study comparing these design choices and their combinations. 
The results show that combining the label-specific prior with residual-space training achieves the lowest errors across the reported validation metrics. 
This suggests that reducing the source–target mismatch and simplifying the transport in residual space leads to more stable learning and more uniform performance across the label space.

\paragraph{Label-specific prior.} Given a particular sample $x_1$ with label y, we denote the label-conditional probability path by $p_t(x|x_1,y)$. 
By integrating out $x_1$, we obtain the marginal conditional path $p_t(x|y)$:
\begin{equation}
p_t(x|y) = \int p_t(x|x_1, y) \, q(x_1|y) \, dx_1.
\label{eq:p_t(x|y)}
\end{equation}
We further define a label-marginal vector field $u_t(x|y)$:
\begin{align}
u_t(x|y) := &\int u_t(x|x_1, y) \, \frac{p_t(x|x_1, y) \, q(x_1|y)}{p_t(x|y)} \, dx_1,
\label{eq:u_t(x|y)}
\end{align}
where $u_t(x|x_1,y)$ generates $p_t(x|x_1,y)$. As formally proven in Appendix \ref{app:theorem}, for any distribution $q(x_1|y)$, the vector field $u_t(x|y)$ defined in \eqref{eq:u_t(x|y)} generates the marginal path $p_t(x|y)$ in \eqref{eq:p_t(x|y)}. The corresponding label-marginal flow matching objective is
\begin{equation}
\mathcal{L}_{\mathrm{FM}}^{y}(\theta)
=
\mathbb{E}_{t\sim\mathcal{U}[0,1],\,q(y),\,p_t(x|y)}
\bigl\|v_\theta(t,x,y)-u_t(x| y)\bigr\|^2. 
\label{eq:FM_y}
\end{equation}
Since the label-marginal vector field \(u_t(x| y)\) is generally
intractable, we instead optimize the \textbf{Label-Conditional Flow Matching (LCFM) objective}:
\begin{equation}
\mathcal{L}_{\text{LCFM}}(\theta) = \mathbb{E}_{t \sim \mathcal{U}[0,1] ,\  q(x_1, y) ,\  p_t(x|x_1,y)} \left\| v_{\theta}(t,x,y) - u_t(x|x_1, y) \right\|^2.
\label{LCFM}
\end{equation}
As shown in Appendix~\ref{app:theorem}, 
\(\mathcal{L}_{\mathrm{FM}}^{y}\) and \(\mathcal{L}_{\mathrm{LCFM}}\) differ
only by a constant independent of \(\theta\), and therefore have identical
gradients with respect to \(\theta\). Thus, optimizing
\(\mathcal{L}_{\mathrm{LCFM}}\) is equivalent, in expectation, to optimizing
the label-marginal flow matching objective for \(u_t(x|y)\). Since this
equivalence holds for any valid probability path, we can construct a
label-specific Gaussian probability path
\begin{equation}
p_t(x|x_1, y) = \mathcal N(x | {\mu}_t(x_1, y), \sigma_t ^2(x_1,y)I),
\end{equation}
where ${\mu}_t(x_1, y)$ and $\sigma_t(x_1,y)$ are explicitly label-dependent, defining a label-specific probability path. 
We design it by satisfying the boundary conditions at $t{=}0$:
\begin{align}
\mu_0(x_1, y)=\mu_0(y), \quad  \sigma_0(x_1,y)=\sigma_0(y), \quad p_0(x|y) = \mathcal N(x | \mu_0(y), \sigma_{0}(y)^2I),
\end{align}
and at $t{=}1$:
\begin{align}
\mu_1(x_1,y)=x_1, \quad \sigma_1(x_1,y)=\sigma_{small}, \quad p_1(x|y) \simeq q(x|y).
\end{align}
For finite \(\sigma_{\rm small}\), this endpoint should be interpreted as a Gaussian-smoothed approximation to \(q(x| y)\), becoming exact in the limit \(\sigma_{\rm small}\to 0\). Instead of integrating the flow directly, we simplify the computation by reparameterizing the probability path in terms of a standard normal variable $z_0 \sim \mathcal{N}(0, I)$ (we refer the reader to Appendix \ref{app:rep} for the step-by-step intermediate derivations of the associated flow and vector fields). 
Under this reparameterization, samples from the conditional path can be written as
\begin{align}
x_t=\mu_t(x_1,y)+\sigma_t(x_1,y)z_0.
\end{align}
Therefore, the LCFM loss can be written as
\begin{align}
\mathcal{L}_{\text{LCFM}}(\theta) = \mathbb{E}_{t \sim \mathcal{U}[0,1] ,\  q(x_1, y) ,\  p(z_0)} 
\left\| v_{\theta}(t,x_t,y) - (\dot\mu_t + \dot\sigma_t z_0) \right\|^2.
\label{loss_LCFM}
\end{align}
In this way, minimizing the LCFM objective provides a tractable training objective for approximating the label-marginal vector field \(u_t(x|y)\), which transports the label-specific Gaussian prior \(p(x_0|y)\) at \(t=0\) toward an approximation of the target distribution \(q(x|y)\) at \(t=1\).

\paragraph{Prior setting for mass maps.}
For our work of generating mass maps, we adopt the following parameterization:
\begin{align}
{\mu}_t(x_1, y) &= t {x}_1 + (1-t) {{\mu}}_0(y), \\
\sigma_t(x_1,y) &\equiv\sigma_t(y)= t\sigma_{1}(y) + (1-t)\sigma_0(y),
\end{align}
$\mu_0(y)$, $\sigma_{0}(y)$, and $\sigma_{1}(y)$ are parameters determined by the conditional distribution $q(x_1|y)$ statistics, i.e.
\begin{align}
\mu_0(y) &= \mathbb{E}[x_1|y], \\
\sigma_0(y) &= \alpha \sqrt{\frac{1}{d}\mathrm{tr}\!\left(\mathrm{Cov}[x_1 |y]\right)}\label{sigma0},\\
\sigma_1(y) &= \beta\, \sigma_0(y),
\label{sigma1}
\end{align}
where $d$ is the dimensionality of the map, $\mu_0(y)$ denotes the mean of the target distribution $q(x_1|y)$, $(\sigma_0(y)/\alpha)^2$ is the average marginal variance, i.e. the average of the diagonal entries of the conditional covariance matrix $\mathrm{Cov}[x_1|y]$, and $\alpha$ and $\beta$ are two hyperparameters. In the theoretical formulation, \(\mu_0(y)\) may be interpreted as a full map-shaped conditional mean vector. In our implementation, motivated by the approximate statistical homogeneity of the maps at fixed cosmology, we approximate \(\mu_0(y)\) by a spatially constant scalar mean and broadcast it to all pixels; see Appendix~\ref{app:network} for details.
In this setting, we define an isotropic Gaussian prior with mean $\mu_0(y)$ and variance $\sigma_{0}^2(y)$.

\paragraph{Improving  interpolation statistics.}
To stabilize conditional flow matching across cosmological labels, we introduce a key reparameterization: per-label mean subtraction. 
For our mass mapping task, we are primarily interested in an emulator that can accurately interpolate across the cosmological parameter space.
Although the label-specific prior provides a more informative initialization, our ablation results on test interpolation points show that using the label-specific prior alone can lead to systematically worse performance across validation metrics (see Fig.~\ref{summary_mean_max_std}).
This indicates that, for our setting, improving interpolation robustness requires not only an informative prior but also an appropriate representation for learning conditional dynamics. Directly training the model in the original coordinate space $x$ forces the network to simultaneously model 
(i) a largely trivial label-dependent translation (mean field shift), and 
(ii) the true residual structure, while the conditional data distributions $q(x_1|y)$ exhibit substantial label-dependent variations in their mean fields, overall amplitudes, and higher-order structure (see Appendix~\ref{app:ablation} for details).
To reduce this burden and stabilize interpolation, we introduce a residual-space reparameterization via per-label mean subtraction
\begin{equation}
r_t = x_t - \mu_0(y),
\end{equation}
which removes the label-dependent mean component from the learning problem and allows the network to focus on modeling the residual structure that is more transferable across nearby cosmologies. The corresponding conditional path in residual space can be sampled through the reparameterization:
\begin{equation}
r_t =\mu_t - \mu_0 + \sigma_t z_0  =t(x_1-\mu_0+(\sigma_1-\sigma_0)z_0)+\sigma_0z_0.
\label{res}
\end{equation}
In this residual space, we have $r_0 \sim \mathcal{N}(0,\sigma_0^2 I)$ and $r_1 \sim \mathcal{N}(x_1 - \mu_0, \sigma_{\mathrm{1}}^2 I)$. Differentiating this path
with respect to time gives the pathwise target velocity:
\begin{align}
&\dot r_t= x_1-\mu_0+(\sigma_1-\sigma_0)z_0.
\end{align}

\noindent
Now we introduce the \textbf{Residual Label-Conditional Flow Matching (RLCFM) objective}:
\begin{equation}
\mathcal{L}_{\text{RLCFM}}(\theta) = \mathbb{E}_{t \sim \mathcal{U}[0,1] ,\  q(x_1, y) ,\  p(z_0)} \left\| v_{\theta}(t,r_t,y) - \dot r_t \right\|^2.
\label{RLCFM}
\end{equation}
At inference time, we generate \(r_1\) by integrating the learned ODE in residual space and reconstruct the corresponding transformed map as \(x_1=r_1+\mu_0(y)\). In Appendix~\ref{app:ablation}, we empirically show that this residual-space formulation improves interpolation robustness across cosmological parameters.
\section{Convergence maps dataset} \label{sec:dataset}

\subsection{Convergence map}
A convergence map represents the two-dimensional projection of the three-dimensional matter perturbations along the line-of-sight, weighted by the lensing efficiency. 
Because it directly traces the projected matter fluctuations, the convergence map is often referred to as a mass map. 
The convergence field can be written as
\begin{equation}
    \kappa(\boldsymbol{\theta}) \;=\; \int_0^{\chi_h} d\chi \, g(\chi)\,\delta(\chi\boldsymbol{\theta}, \chi),
\end{equation}
where $\chi$ is the comoving distance, $\chi_h$ is the comoving distance to the horizon,
$\delta$ is the matter density contrast and $g(\chi)$ is the lensing efficiency, which serves as an ``average'' weight function
\begin{equation}
g(\chi)=\frac{3H_0^2\Omega_m}{2c^2} \frac{\chi}{a(\chi)}\int_{\chi}^{\chi_h} d\chi' \frac{n_s(z)}{\bar{n}_s}\frac{dz}{d\chi'}\frac{\chi'-\chi}{\chi'},
\end{equation}
where $n_s(z)$ is the redshift distribution of the source galaxies and $\bar{n}_s$ is their mean number density. 
The lensing efficiency thus depends on the source galaxy redshift distribution $n_s(z)$.
In this work, we adopt the non-tomographic source redshift distribution, which yields a single source-averaged lensing efficiency $g(\chi)$ for the full source sample.

\subsection{Dataset}
\begin{figure}
  \centering
  \begin{minipage}{0.45\textwidth}
    \includegraphics[scale=0.45]{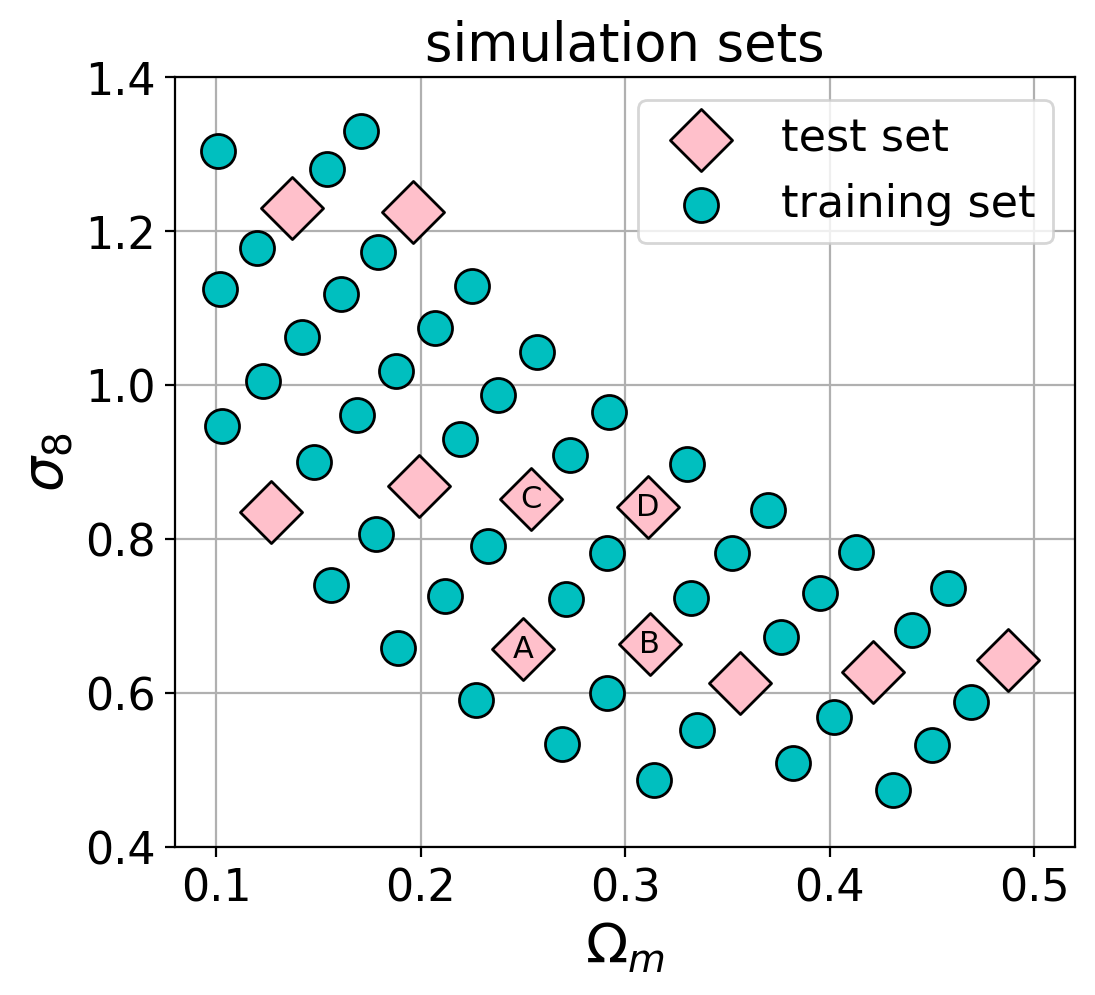} 
  \end{minipage}
  \hfill
  \begin{minipage}{0.54\textwidth}
    \caption{Cosmological parameter grid from \cite{perraudin2021emulation}. The diamonds and circles denote the 11 test sets and 46 training sets, respectively. The cosmologies labeled A, B, C, and D are examined in more detail in Section~\ref{sec:results}.}
    \label{figure_grid_test_train}
  \end{minipage}
\end{figure}

The convergence map dataset employed in this work was introduced in~\cite{fluri2019cosmological, perraudin2021emulation} and generated from N-body simulations with ray-tracing.
It consists of 57 different cosmological models within the standard flat $\Lambda$CDM framework. 
Each model is specified by a pair of parameters, $\Omega_m$ and $\sigma_8$, and provides 12,000 maps, yielding a total of 684,000 samples. The convergence maps are represented as $128{\times}128$ pixel maps, each covering a $5^\circ{\times}5^\circ$ patch of the sky. In this work, we select 46 cosmologies (552,000 samples) as training set and 11 cosmologies (132,000 samples) as the test set \cite{perraudin2021emulation}, as depicted in Fig.~\ref{figure_grid_test_train}.

\paragraph{Data pre-processing.}
Applying suitable reversible transformations to the original data can
make the generative model easier to train. The convergence field is well
approximated by a lognormal random field
\cite{taruya2002lognormal, clerkin2017testing, boruah2022map},
which we parametrize as $\kappa = e^{x} - \kappa_{\min}$
where $x$ is a Gaussian random field. 
In the dataset used here the
stored maps already contain the shifted field $\kappa + \kappa_{\min}$,
i.e. the pixel values correspond to $\kappa_{\rm shifted}\equiv\kappa+\kappa_{\min}=e^x$.
We therefore apply an element-wise logarithm to the stored maps and use the resulting field \(x=\log \kappa_{\rm shifted}\) as the target for training.
Next, we standardize the data by subtracting the global mean $\mu_g$ and dividing
by the global standard deviation $\sigma_g$ of all pixels across the training maps,
\begin{align}
\tilde{x}
= \frac{x - \mu_g}{\sigma_g}.
\label{g}
\end{align}
Then we subtract the mean to get a residual variable $r=\tilde{x}-\mu_0$, as depicted in Section~\ref{2.3}.
We train our RLCFM model to predict the approximated Gaussian field $r$ in this transformed space, and at inference time we map the generated samples back to the shifted convergence field by inverting these transformations.

\section{Quantitative comparison metrics with summary statistics} \label{sec:metrics}

To evaluate how well the RLCFM-generated mass maps reproduce the
statistical properties of the N-body simulations, we combine
cosmology-oriented summary statistics with computer-vision similarity
metrics. 
We adopt the same statistical evaluation procedure as in
\cite{perraudin2021emulation}, thereby assessing both the visual agreement and the physical consistency of the generated maps.

\paragraph{1. Pixel and peak value distributions.}
We first consider one-point statistics of the convergence field. 
The pixel distribution is characterized by the histogram of pixel values,
$N_{\mathrm{pixels}}(\kappa)$, binned in the threshold variable
$\kappa$. 
This statistic traces the marginal probability distribution function of the field and is sensitive to departures from Gaussianity, such as skewness and heavy tails induced by non-linear structure formation. 
In addition, we measure the distribution of local maxima, $N_{\mathrm{peak}}(\kappa)$, defined as pixels whose values exceed those of their neighbors. 
The peak histogram directly probes the abundance of high $\kappa$ features associated with massive halos and projected large-scale structures. 
Together, $N_{\mathrm{pixels}}$ and $N_{\mathrm{peak}}$ provide a compact yet informative characterization of the one-point properties of the field and its peak population, capturing both the Gaussian baseline and simple non-Gaussian deviations in the amplitude of convergence fluctuations.

\paragraph{2. Angular power spectrum $C_\ell$.}
We characterize the two-point statistics of the convergence field by
its angular power spectrum $C_\ell$. 
The quantity $C_\ell$ encodes the variance of Fourier modes as a function of multipole $\ell$ and thus describes the amplitude of fluctuations across angular scales. 
As a standard observable in large-scale structure and weak-lensing analyses, it provides a compact description of the field in the Gaussian regime and is closely connected to the underlying matter power spectrum and the growth of structure. 
In our context, matching $C_\ell$ between simulated and generated mass maps ensures that the two-point statistics and scale-dependent variance are correctly reproduced.

\paragraph{3. Bispectrum $B_{\ell}$.}
As a three-point statistic, the bispectrum probes higher-order correlations of the convergence field and is intrinsically sensitive to its non-Gaussian structure. 
While the power spectrum $C_\ell$ fully describes a Gaussian field, it cannot capture phase correlations or the mode coupling generated by non-linear structure formation. 
The bispectrum, defined for triplets of modes forming a closed triangle in multipole space, directly quantifies these effects and is particularly sensitive to the presence of massive halos, filamentary structures, and other non-linear features. 
As such, it provides information that is complementary to $C_\ell$ and enhances the discriminating power between different cosmological models and between simulated and generated mass maps. 

\paragraph{4. Minkowski functionals $V_0,V_1,V_2$.}
To quantify the morphology of the convergence fields, we measure the Minkowski functionals on thresholded maps. 
For a given threshold $\nu$, we define the excursion set $Q_\nu = \{\mathbf{x} \mid \kappa(\mathbf{x}) \ge \nu\}$. 
In two dimensions, this set is fully characterized by three Minkowski functionals $(V_0, V_1, V_2)$
\begin{align}
V_0(\nu) = \frac{1}{A} \int_{Q_\nu} dA, 
\quad
V_1(\nu) = \frac{1}{4A} \int_{\partial Q_\nu} ds,
\quad
V_2(\nu) = \frac{1}{2\pi A} \int_{\partial Q_\nu} \kappa_g \, ds,
\end{align}
where $A$ is the total area of the map, $\partial Q_\nu$ its boundary, and $\kappa_g$ the geodesic curvature. 
Intuitively, $V_0$ describes the area fraction above the threshold, $V_1$ the total boundary length of the excursion set, and $V_2$ the Euler characteristic, i.e. the number of connected components minus the number of holes. 
These statistics provide a compact characterization of the geometric and topological properties of the convergence field and are sensitive to its non-Gaussian features.

\paragraph{5. Power spectrum correlation matrices $R_{\ell\ell'}$.}
We estimate the covariance of the binned power spectrum and its normalized Pearson correlation matrix $R_{\ell\ell'}$ in order to characterize the cross-scale coupling of $C_\ell$. 
The correlation matrix removes the overall amplitude of the covariance and isolates the pattern of linear dependence between different multipoles, thereby encoding how fluctuations on different angular scales are correlated within the ensemble. 
In our context, $R_{\ell\ell'}$ is particularly useful to assess whether the RLCFM model reproduces the mode coupling and sample variance structure of the N-body simulations: significant deviations of the RLCFM correlation matrix from the N-body reference would indicate mismatches in higher-order statistics, even if the mean power spectrum is well reproduced.
In addition, we summarize its overall correlation strength using the Frobenius norm.

\paragraph{6. Multi-scale structural similarity (MS-SSIM).}
MS-SSIM \cite{wang2004image,wang2003multiscale} provides a measure of image similarity that is sensitive to structure across multiple spatial scales. Unlike one-point statistics or purely correlation-based measures, it captures how local contrast, luminance and morphological features are arranged in the maps, and is therefore closely related to the overall appearance of sharpness, filamentarity and halo-like peaks. 
In our analysis, MS-SSIM quantifies to what extent the RLCFM-generated mass maps preserve the small and large-scale spatial organization seen in the N-body simulations, and thus complements the cosmological summary statistics by probing the structural realism of the samples.

We quantify the agreement between RLCFM and N-body pixel and peak distributions using the Wasserstein\mbox{-}1 distance $W_1(P,Q)$, defined as the minimal average transport cost between two probability distributions. 
Since $W_1$ is sensitive to the overall scale of the variables, we first standardize the pixel values by subtracting the mean and dividing by the standard deviation measured from the N-body maps at the corresponding cosmology, and apply the same transformation to the RLCFM samples. 
We then evaluate the Wasserstein\mbox{-}1 distance on these standardized variables, yielding a normalized $W_1$. 

For each statistic $S(x)$ (e.g., power spectrum $C_\ell$, bispectrum $B_\ell$ and Minkowski functionals $V_{0,1,2}$), we compute the sample mean over the RLCFM-generated maps and over the N-body maps. 
On a fixed evaluation range $x \in[x_{\min},x_{\max}]$, we then define the bin-wise fractional difference
\begin{equation}
f_S
\equiv \Big\langle \,\Big|\, 
\frac{ S_{\mathrm{RLCFM}}(x)- S_{\mathrm{N\text{-}body}}(x)}
     {S_{\mathrm{N\text{-}body}}(x)} 
\,\Big|\, \Big\rangle_{x\in[x_{\min},x_{\max}]},
\label{fs}
\end{equation}
where $\langle \cdot \rangle_{x\in[x_{\min},x_{\max}]}$ denotes an average over bins within the evaluation range, so that $f_S$ summarizes the typical relative error between the RLCFM emulator and the N-body reference for the statistic $S(x)$.

To summarize the overall strength of the correlation matrices, we rely on the Frobenius norm $\|R\|_F \equiv \left( \sum_{i,j} |r_{ij}|^2 \right)^{1/2}$ as a scalar summary.  
Accordingly, we report the fractional contrast between the norms of the RLCFM and N-body correlation matrices as
\begin{equation}
f_R \equiv \Big| \frac{\|R_{\rm RLCFM}\|_{F}-\|R_{\rm N\text{-}body}\|_{F}}{\|R_{\rm N\text{-}body}\|_{F}} \Big|.
\label{fr}
\end{equation}

The Multi-Scale Structural Similarity Index (MS-SSIM) provides a means to probe mode collapse, i.e. situations in which the generator concentrates on a limited subset of the training distribution.
Since conventional summary statistics can remain unchanged under such failures, MS-SSIM provides a complementary measure of structural similarity across many image pairs. 
It returns a score in $[0,1]$ for any two inputs (1 for identical and 0 for maximally dissimilar images). Focusing on ensemble-level behavior, and following the terminology of Ref. \cite{perraudin2021emulation}, we quantify the significance of the difference in average scores via
\begin{equation}
s_{\rm SSIM} \equiv
\frac{\langle {\rm SSIM}_{\rm RLCFM} \rangle - \langle {\rm SSIM}_{\rm N\text{-}body} \rangle}
{\big(\sigma[{\rm SSIM}_{\rm RLCFM}] + \sigma[{\rm SSIM}_{\rm N\text{-}body}]\big)/2},
\label{ssim}
\end{equation}
where $\langle\cdot\rangle$ denotes the sample mean and $\sigma[\cdot]$ the sample standard deviation. Large values of $|s_{\rm SSIM}|$ indicate a significant discrepancy between the two ensembles, while small values suggest comparable structural variability.
\section{Results} \label{sec:results}
We trained the RLCFM model described in Appendix~\ref{app:network} and it took us about 7 hours to train the model for 65,000 steps (about 15 epochs) on an NVIDIA A100 Tensor Core GPU. Unless otherwise stated, all samples and statistics reported below are generated using the checkpoint at 50{,}000 training steps. For each pair of labels, we generated 5,000 RLCFM maps and randomly sampled 5,000 N-body maps for statistical comparisons. In this section, we first show the maps generated by the RLCFM model. We then present a series of statistical comparisons for labels A, B, C, and D on the test set shown in Fig.~\ref{figure_grid_test_train}, and finally, we present the statistical results for all labels on both the training and test sets.

\subsection{Visual comparison}
\begin{figure}[!htbp]  
  \centering
  \includegraphics[width=\linewidth]{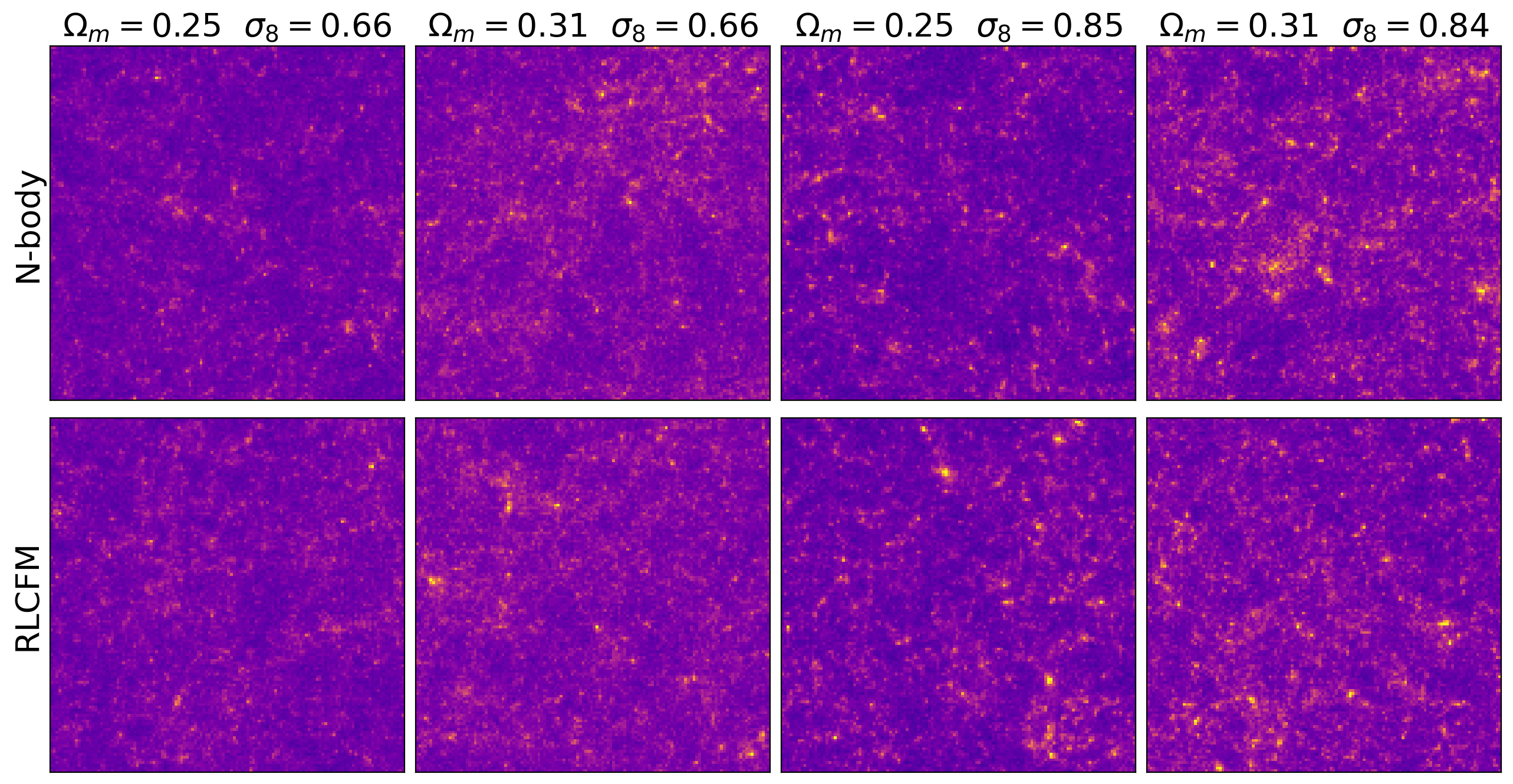}  
  \caption{Representative N-body maps and RLCFM-generated maps for the four representative cosmological parameter sets A, B, C, and D in Fig.~\ref{figure_grid_test_train}. The generated maps are in very good visual agreement with the N-body references, with closely matched large-scale morphology and small-scale structures across the different cosmologies.}  \label{figure_maps_nbody_vs_fm}
\end{figure}

Fig.~\ref{figure_maps_nbody_vs_fm} presents example mass maps from the RLCFM model alongside the
representative N-body maps for the A, B, C and D choices of
$(\Omega_m,\sigma_8)$. The generated and original maps are visually
almost indistinguishable, and the morphology of the structures evolves
in a consistent way as the cosmological parameters are varied. In
particular, as expected from theory, increasing $\Omega_m$ enhances the
overall mass content of the maps, while larger $\sigma_8$ leads to
maps with a higher variance in pixel intensities.
\begin{figure}
  \centering  \includegraphics[width=0.7\linewidth]{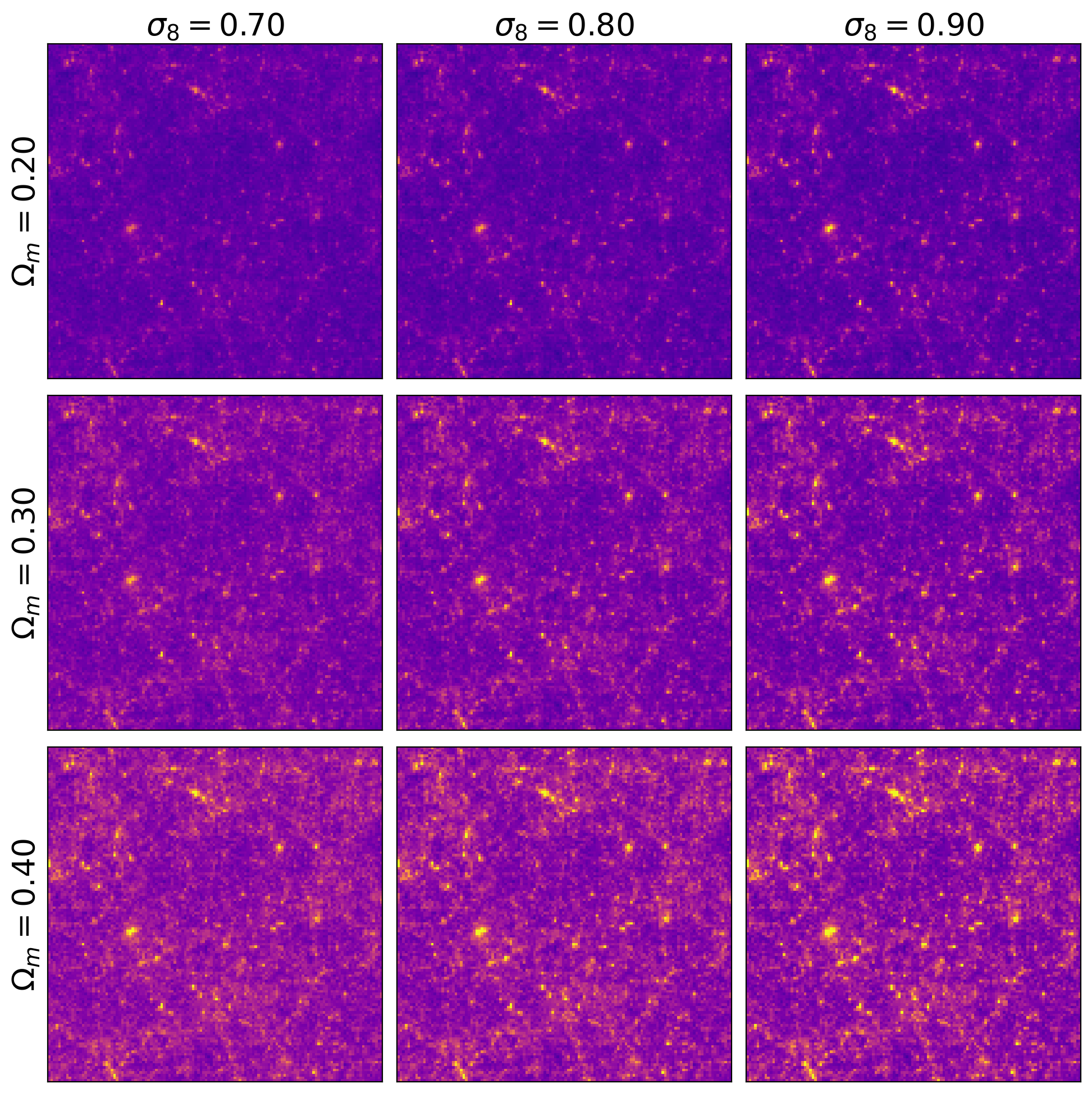} 
  \caption{Maps generated from the same initial noise under different cosmological parameters $(\Omega_m,\sigma_8)$. The smooth changes across the grid indicate that the model has learned a continuous parameter dependence. As expected from theory, increasing $\Omega_m$ enhances the overall mass content and raises the mean convergence level, while increasing $\sigma_8$ produces stronger clustering and larger small-scale fluctuations in the maps.}  \label{figure_cosmo_grid_FM}
\end{figure}
Fig.~\ref{figure_cosmo_grid_FM} shows mass maps generated by the conditional model for a
grid of cosmological parameters. The smooth transition from low to high
mass density across the grid demonstrates that the model has learned a
continuous mapping from $(\Omega_m,\sigma_8)$ to the corresponding mass
distributions: increasing $\Omega_m$ raises the overall convergence
level, while larger $\sigma_8$ produces more pronounced small-scale
fluctuations in the maps.

\subsection{Summary statistics}
\begin{figure}
  \centering
  \includegraphics[width=\linewidth]{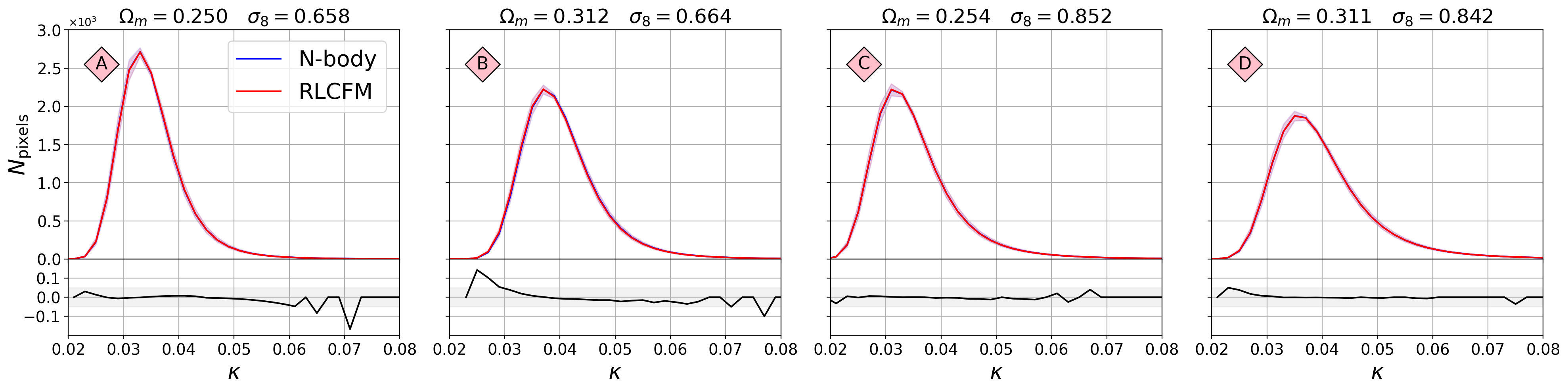}
  \includegraphics[width=\linewidth]{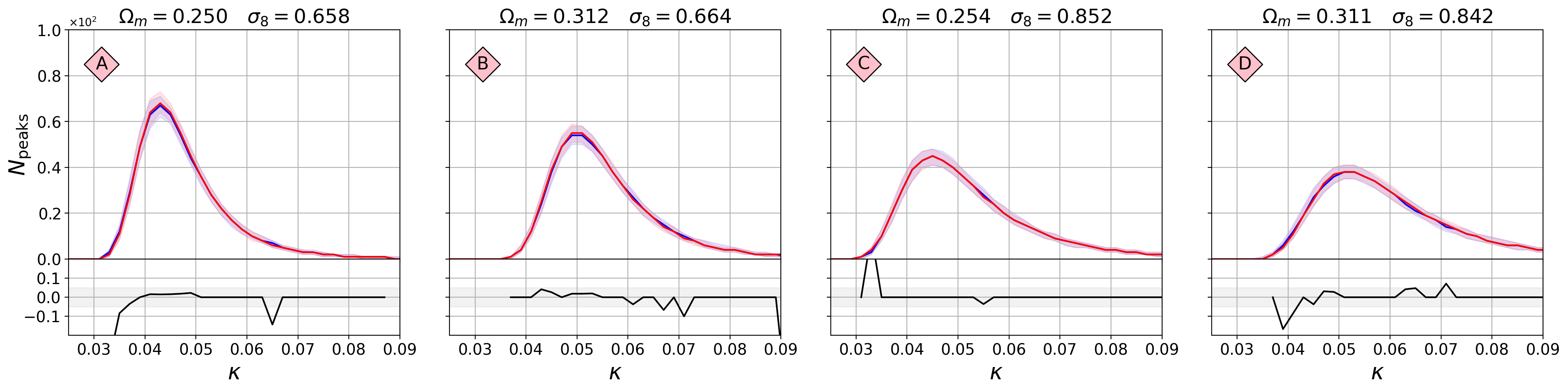}
  \caption{Comparison of pixel histograms (top) and peak histograms (bottom) between N-body and RLCFM-generated mass maps for models A, B, C, and D. The solid line corresponds to the median histogram, and the bands correspond to the 32\% and 68\% percentiles of the ensemble of histograms. The lower panel in each subplot shows the fractional difference of the median histograms.}
  \label{pixel/peak}
\end{figure}

Fig.~\ref{pixel/peak} shows the histograms of pixels (top) and peaks (bottom) for RLCFM-generated maps (red) and N\text{-}body maps (blue) for the four models A, B, C, and D. Peaks are defined as pixels whose values exceed all other pixels within a $5\times5$ neighborhood (i.e., the 24 surrounding pixels). The solid line corresponds to the median of the histogram, and the bands correspond to the 32\% and 68\% percentiles. The lower
panel in each subplot shows the fractional difference of the median histograms, $f_{median} = (x_{FM}-x_{N-body})/x_{N-body}$. The x-axis $\kappa$ represents the pixel intensity. We compared the normalized Wasserstein-1 distances for pixel distribution in $ \kappa\in[0.02, 0.08]$, and peak distribution in $ \kappa\in[0.025, 0.09]$. For the pixel value distribution, the distances are $W_{pixel}=0.004, 0.012, 0.003, 0.002$ for models A, B, C, and D, respectively. For the peak value, the distances are $W_{peak}=0.014, 0.012, 0.008, 0.008$. 
These results indicate that the normalized Wasserstein-1 distances between the RLCFM and N-body pixel and peak histograms are below 0.02 over the evaluated ranges.

\begin{figure}
  \centering
  \includegraphics[width=\linewidth]{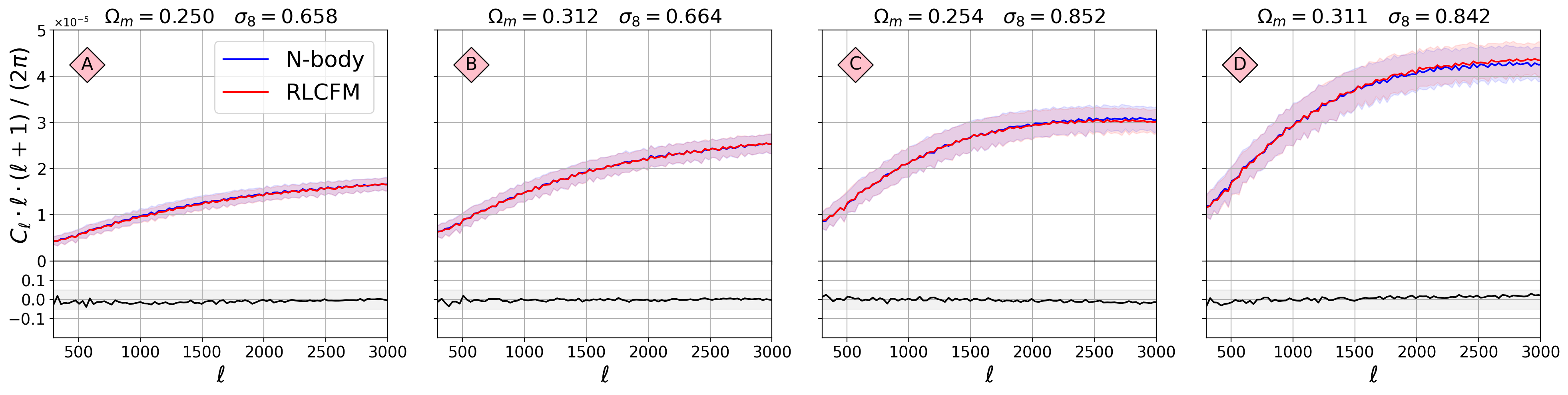}
  \includegraphics[width=\linewidth]{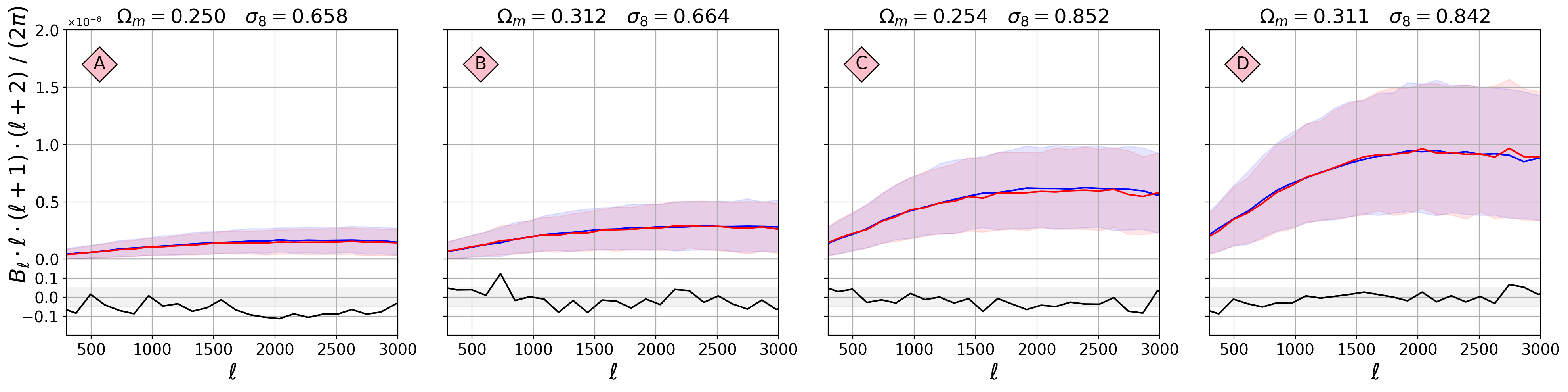}
  \caption{Comparison of power spectra (top) and bispectra (bottom) between N-body and RLCFM-generated mass maps for models A, B, C, and D. The structure of this figure is the same as for Fig.~\ref{pixel/peak}.}
  \label{2pt/3pt}
\end{figure}
The power spectra and bispectra are evaluated using \textsc{LensTools}~\cite{petri2016mocking}. For the bispectrum, we adopt the folded configuration, using the default ratio of 0.5 between one of the triangle sides and the base. Fig.~\ref{2pt/3pt} shows the comparison of the two-point (top) and three-point (bottom) statistics. We bin the multipole range $\ell \in [200, 6000]$ into 200 bins for the power spectrum and 50 bins for the bispectrum, and evaluate the difference defined in Eq.~\eqref{fs} over $\ell \in [300, 3000]$ in both cases. For the two-point statistics, the differences are $f_{C_\ell}=0.014, 0.005, 0.008, 0.012$ for models A-D, all below $2\%$. For the three-point statistics, we obtain $f_{B_\ell}=0.077, 0.031, 0.027, 0.032$, indicating similarly good agreement. These results demonstrate that the RLCFM model successfully reproduces both the Gaussian and non-Gaussian information encoded in the mass field. 
\begin{figure}
  \centering
  \includegraphics[width=\linewidth]{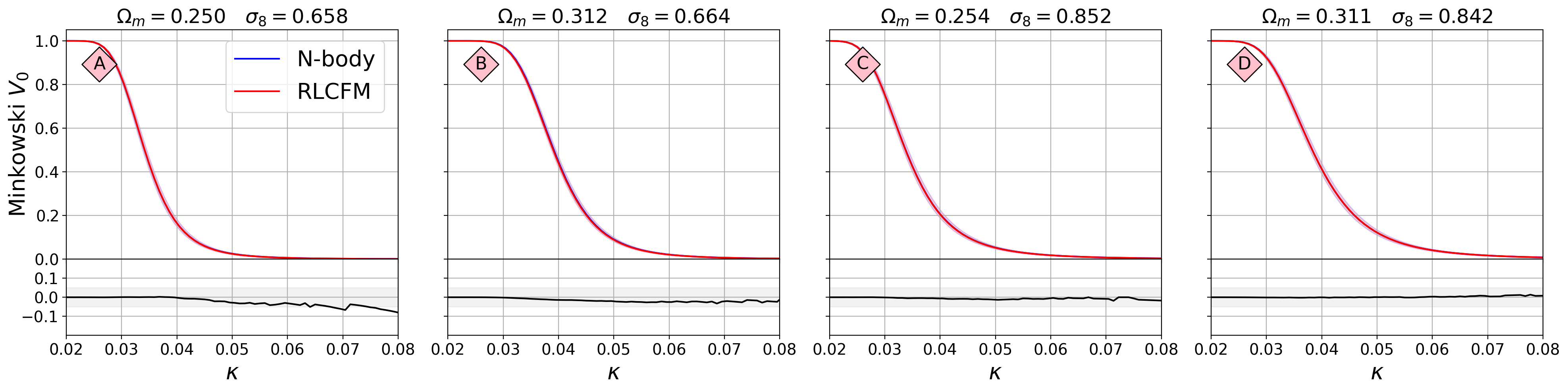}
  \includegraphics[width=\linewidth]{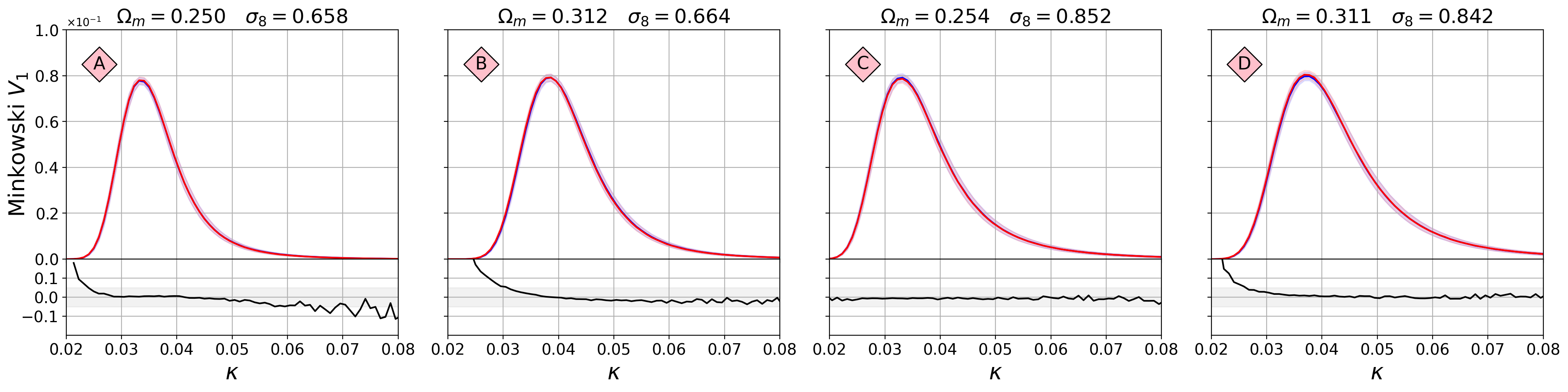}
  \includegraphics[width=\linewidth]{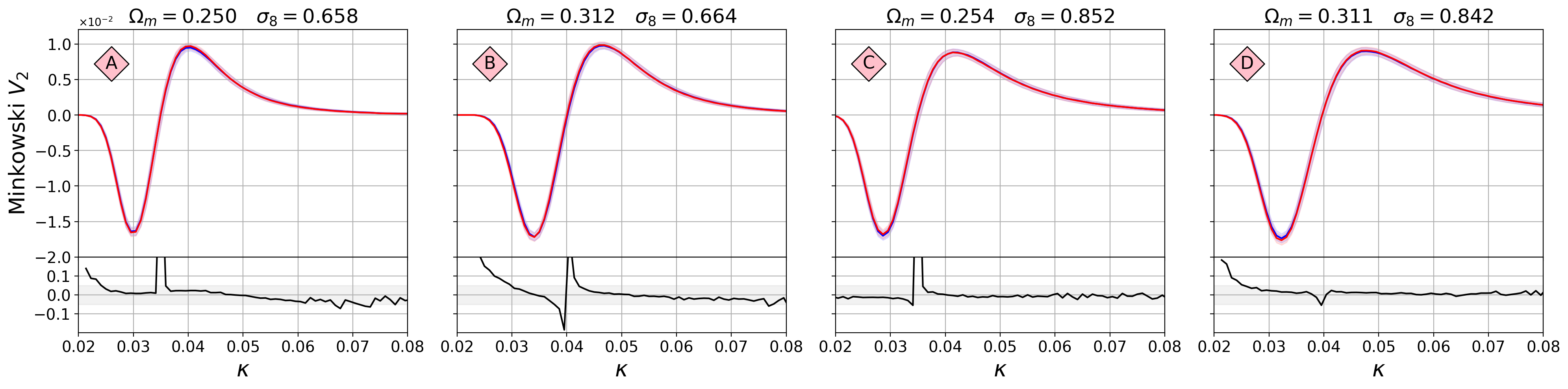}
  \caption{Comparison of the Minkowski functionals $V_0$ (top), $V_1$ (middle), and $V_2$ (bottom) between N-body and RLCFM-generated mass maps for models A, B, C, and D. The structure of this figure is the same as for Fig.~\ref{pixel/peak}.
  The large differences in the median fractional difference plots are due to numerical instability close to the value of $V=0$.}
  \label{V}
\end{figure}

The Minkowski functionals are shown in Fig.~\ref{V} and are also computed with \textsc{LensTools}~\cite{petri2016mocking}.
The x-axis shows the threshold $\kappa$, above which the excursion set (the “islands”) is defined. For each threshold, $V_0$ (top row) measures the area of these islands, $V_1$ (middle row) their total boundary length, and $V_2$ (bottom row) their Euler characteristic. We evaluate $V_{0,1,2}$ at 100 thresholds uniformly spaced in $\kappa \in [0.01, 0.1]$, and compute the difference over the interval $\kappa \in [0.02, 0.08]$.
The large variations in the median fractional difference plots arise from numerical instability near $V=0$. Since the relative difference becomes meaningless in this regime, for $f_{V_{1,2}}$ we restrict its evaluation to $|V_{1,2}|>10^{-5}$.  The resulting values are $f_{V_0} = 0.027, 0.014, 0.006, 0.002$, $f_{V_1} = 0.031, 0.031, 0.007, 0.019$, and $f_{V_2} = 0.030, 0.039, 0.072, 0.022$. These small deviations indicate that the RLCFM-generated maps successfully reproduce the morphological and topological information encoded in the Minkowski functionals of the N-body mass fields.

\begin{figure}
  \centering
  \includegraphics[width=\linewidth]{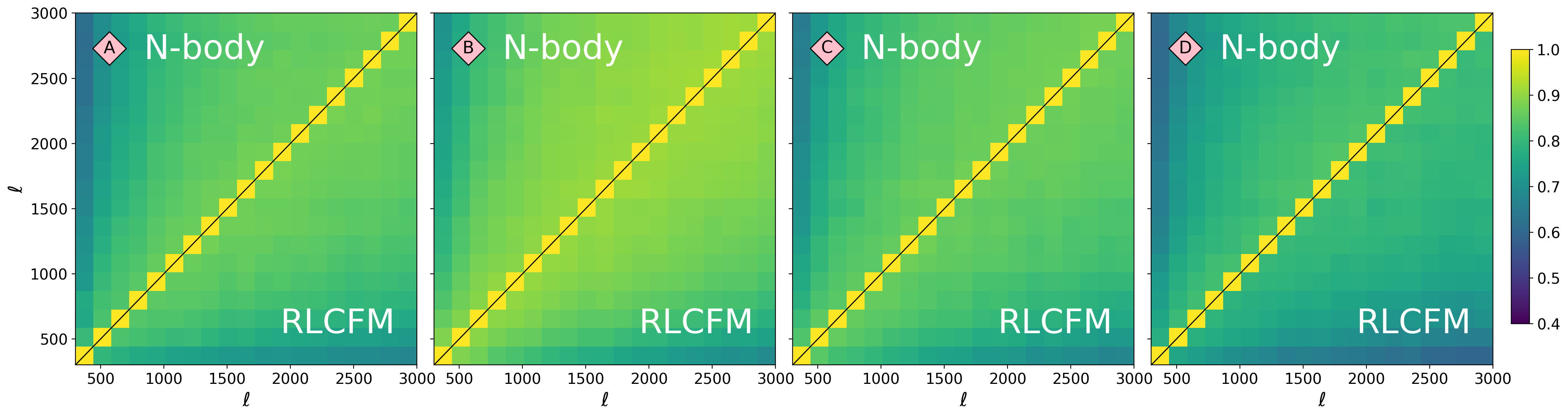} 
  \caption{Pearson’s correlation matrices for models A, B, C, and D. In each matrix, the upper triangular part shows the correlations from the N-body realizations, while the lower triangular part shows those from the RLCFM samples.}
  \label{corr}
\end{figure}
The Pearson correlation matrices are shown in Fig.~\ref{corr}. The upper and lower triangular parts of each matrix correspond to the correlations measured from the N-body realizations and from the RLCFM samples. We summarize the overall correlation strength over the range $\ell \in [300, 3000]$ by comparing the Frobenius norms of the RLCFM and N-body correlation matrices using Eq.~(\ref{fr}). The resulting values are $f_R=0.037,0.023,0.002,0.013$, indicating a small contrast in the total Frobenius-norm amplitude of the correlation matrices.

We calculate the MS-SSIM between the RLCFM and N-body maps, then evaluate the significance of the difference in average scores via Eq.~(\ref{ssim}). The significance values for models A, B, C and D are: $s_{SSIM}=0.023, -0.159, 0.205, -0.092$. This indicates very good agreement in the MS-SSIM statistic for these models.

\begin{figure}
  \centering
  \includegraphics[width=\linewidth]{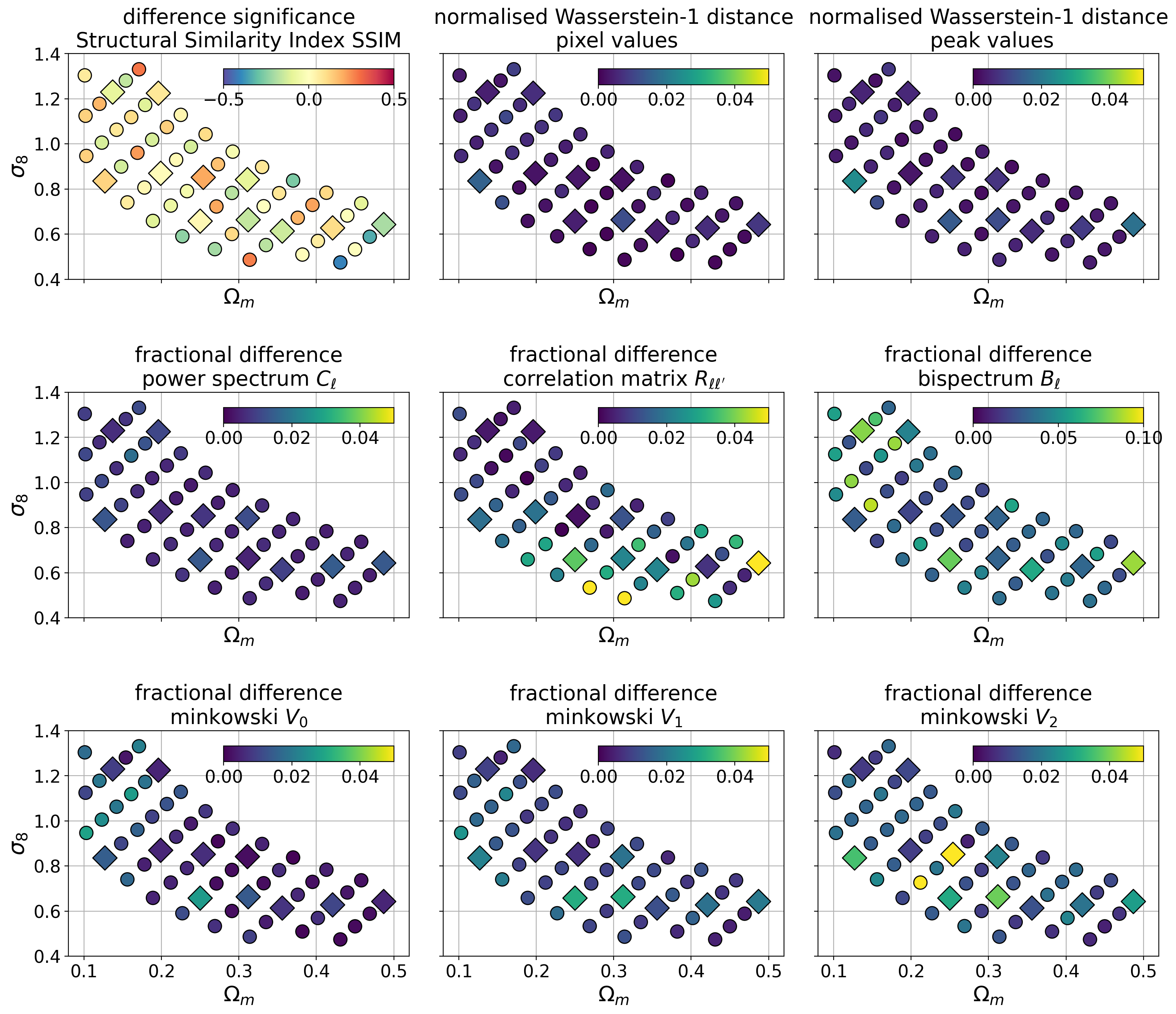} 
  \caption{Differences between the summary statistics measured from the N-body and RLCFM-generated maps over the $(\Omega_m,\sigma_8)$ grid. The panels show the SSIM significance, the normalized Wasserstein-1 distances for the pixel-value and peak distributions, the mean absolute fractional differences for the power spectrum and bispectrum, the fractional difference in the Frobenius norm of the power spectrum correlation matrix, and the mean absolute fractional differences for the three Minkowski functionals $V_0$, $V_1$, and $V_2$. Circles correspond to training cosmologies and diamonds to test cosmologies. The color indicates the magnitude of the discrepancy for each cosmological parameter point. The agreement is generally very good across the grid, with only slightly enhanced discrepancies in a small number of edge and test cosmologies.}
  \label{overview}
\end{figure}

Finally, using the same grids of training and test cosmologies shown in Fig.~\ref{figure_grid_test_train}, we examine how the different summary statistics vary across the cosmological parameter space. Figure~\ref{overview} displays the resulting nine summary measures over the $(\Omega_m,\sigma_8)$ plane: (i) top left: significance of the difference $s_{\mathrm{SSIM}}$ in the multi--scale structural similarity index between RLCFM and N-body maps; (ii) top center: normalized Wasserstein-1 distance of the pixel value distributions, $W^{\mathrm{pixel}}_{1}$; (iii) top right: normalized Wasserstein-1 distance of the peak value distributions, $W^{\mathrm{peak}}_{1}$; (iv) middle left: mean absolute fractional difference of the power spectrum, $f_{C_\ell}$; (v) middle center: fractional difference in the Frobenius norm of the power spectrum correlation matrices, $f_{R}$; (vi) middle right: mean absolute fractional difference of the bispectrum, $f_{B_\ell}$; (vii) bottom left: mean absolute fractional difference of the Minkowski functional $V_0$, $f_{V_0}$; (viii) bottom center: mean absolute fractional difference of the Minkowski functional $V_1$, $f_{V_1}$; and (ix) bottom right: mean absolute fractional difference of the Minkowski functional $V_2$, $f_{V_2}$.
In all panels, circles denote cosmologies from the training set and diamonds
denote cosmologies from the test set, while the color encodes the value of the
corresponding statistic for a given $(\Omega_m,\sigma_8)$. Averaging separately over the training and test cosmologies, we obtain mean absolute SSIM significances of 0.131 and 0.111, respectively. The mean normalized Wasserstein-1 distances are 0.004 and 0.006 for the pixel-value distribution and 0.004 and 0.010 for the peak distribution. The mean discrepancies are 0.006 and 0.010 for the power spectrum, 0.017 and 0.019 for the Frobenius-norm contrast of the power spectrum correlation matrices, and 0.039 and 0.047 for the bispectrum, while for the Minkowski functionals they are 0.009 and 0.010 for $V_0$, 0.011 and 0.017 for $V_1$, and 0.014 and 0.026 for $V_2$.

Overall, the agreement between the RLCFM maps and the N-body simulations is very good over most of the grid. The Wasserstein distance of the pixel histograms is small and almost constant (typically $W^{\mathrm{pixel}}_{1}<0.01$)\footnote{Here and in the following, 
“typically” refers to the 80th percentile over the 57 cosmological 
parameter points.}, indicating that the one–point distributions are reproduced accurately for all cosmologies. The peak histogram (typically $W^{\mathrm{peak}}_{1}<0.01$), power spectrum (typically $f_{C_\ell}<1\%$), and Minkowski functional (typically $f_{V_0}<1\%$, $f_{V_1}<2\%$, $f_{V_2}<3\%$) fractional differences are at the level of only a few percent, with the smallest discrepancies found near the center of the parameter grid, where neighboring cosmologies also show very similar values of these statistics. Slightly larger deviations appear towards the edges of the grid: in particular, the correlation matrix statistic (typically $f_{R}<3\%$) and the SSIM significance show enhanced differences for high $\Omega_m$ and low $\sigma_8$, whereas the bispectrum fractional difference (typically $f_{B_\ell}<5\%$) peaks at low $\Omega_m$ and high $\sigma_8$. We further find that the magnitude of the statistical differences between the RLCFM maps and the N-body simulations is very stable across the cosmological parameter space: at each cosmological parameter point the values of all statistics for the training and test maps are nearly identical, with the test and edge cosmologies showing only slightly larger discrepancies on average, and the differences vary smoothly from one cosmology to the next, indicating that the model generalizes well across the sampled parameter grid.

\subsection{Distribution-level comparison of summary statistics}

We compare a distribution of summary statistics calculated from: 
(1) N-body maps generated from random initial conditions and 
(2) emulator maps generated from random noise at $t{=}0$.
This comparison enables us to assess whether the emulator reproduces the distributional properties of cosmological summary statistics and their realization-to-realization variability.

\begin{figure}
  \centering
  \includegraphics[width=\linewidth]{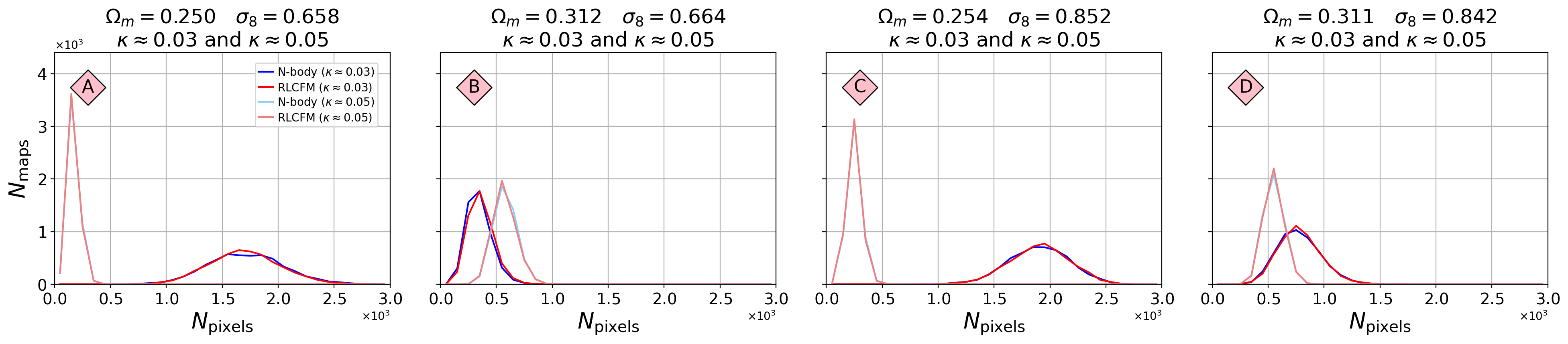}
  \includegraphics[width=\linewidth]{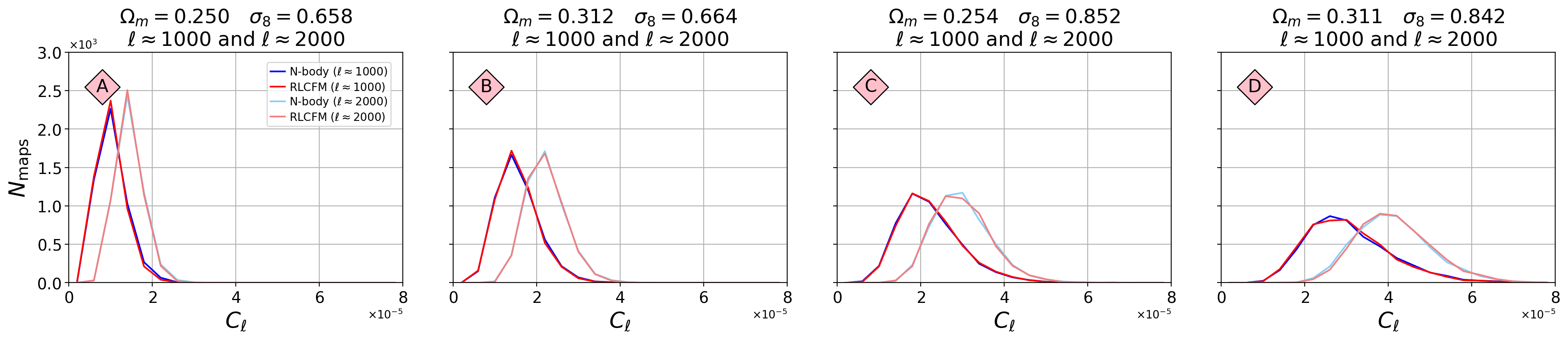}
  \includegraphics[width=\linewidth]{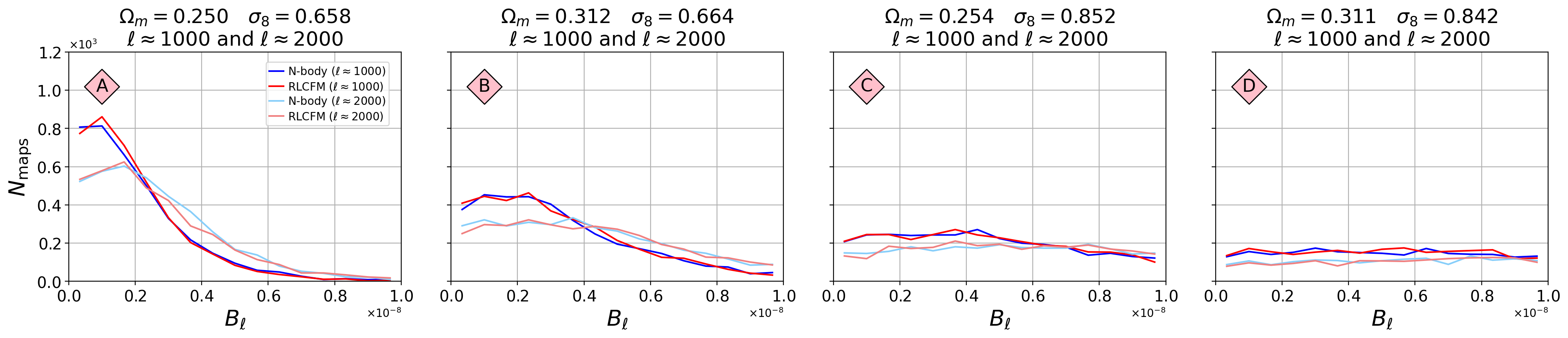}
  \caption{Distributions at representative bins of pixel-value, power spectrum, and bispectrum quantities for the four cosmologies A, B, C, and D. Top: distribution of map counts as a function of the number of pixels per map falling within representative pixel-value bins centered around $\kappa \approx 0.03$ and $\kappa \approx 0.05$. Middle and bottom: distributions of the corresponding power spectrum and bispectrum values for representative multipole bins around $\ell \approx 1000$ and $\ell \approx 2000$. In each panel, the blue (light blue) and red (light red) curves denote the N-body and RLCFM results, respectively.}
  \label{bin_wise}
\end{figure}
\begin{figure}[!htbp]  
  \centering 
  \includegraphics[width=\linewidth]{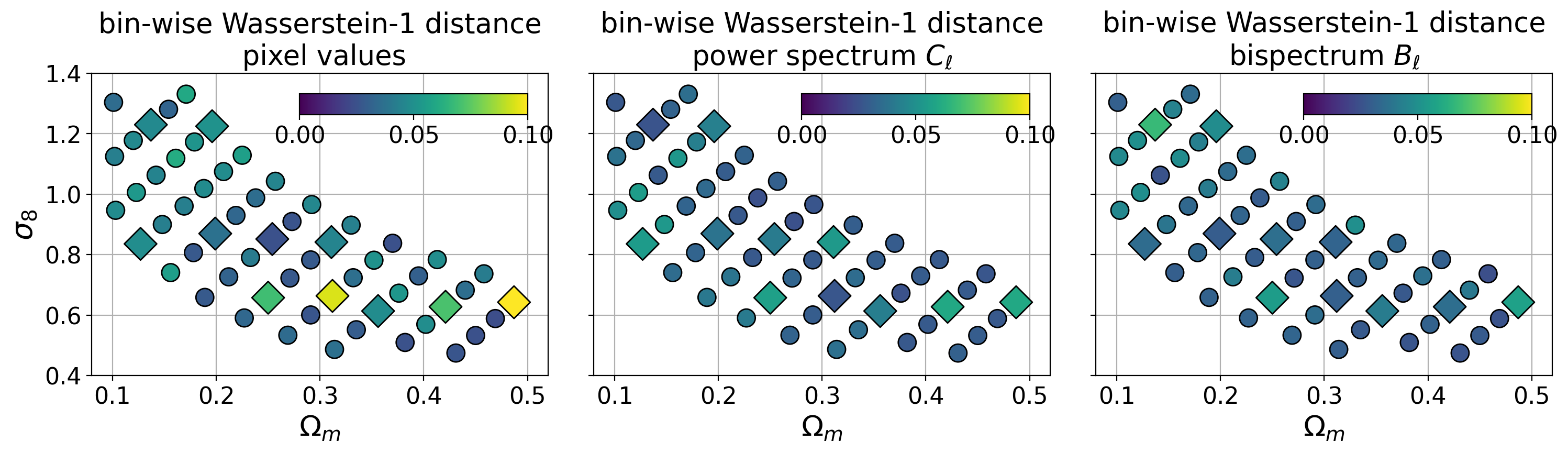}
  \caption{Average bin-wise normalized Wasserstein-1 distances between the N-body and RLCFM distributions across the cosmological grid for the pixel statistic (left), power spectrum $C_\ell$ (middle), and bispectrum $B_\ell$ (right). For each cosmology, the scalar value shown is obtained by computing the normalized Wasserstein-1 distance in every individual $\kappa$ or $\ell$ bin and then averaging over all bins.}
  \label{overview_bin_wise}
\end{figure}

We compare the distributions of representative pixel-count, power spectrum, and bispectrum quantities for cosmologies A, B, C, and D, using the corresponding N-body and RLCFM results. 
As shown in Fig.~\ref{bin_wise}, for the pixel statistic we fix a representative pixel-value bin around $\kappa \approx 0.03$ and $\kappa \approx 0.05$, count for each map the number of pixels falling into this bin, and then construct the distribution of map counts as a function of this pixel number. 
For the power spectrum and bispectrum, we instead fix a representative multipole bin around $\ell \approx 1000$ and $\ell \approx 2000$, compute the corresponding $C_\ell$ and $B_\ell$ values for each map, and then obtain the distributions of $C_\ell$ and $B_\ell$ across the ensemble of maps. 
For all three representative quantities, the N-body and RLCFM distributions are broadly consistent for the four test cosmologies. 
In particular, the peak locations and overall widths of the distributions are generally well matched, indicating that the model reproduces not only the typical values of these statistics but also their map-to-map variability. 

To summarize the bin-wise difference over the full cosmological grid, we compute the normalized Wasserstein-1 distance, as defined in Section~\ref{sec:metrics}, between the N-body and RLCFM distributions for each individual \(\kappa\) or \(\ell\) bin. 
We then average these bin-wise Wasserstein-1 distances over the full set of bins, thereby obtaining a single scalar measure for each cosmology and for each summary statistic, as shown in Fig.~\ref{overview_bin_wise}. 
The mean values over the training and test cosmologies are 0.039 and 0.058 for the pixel statistic, 0.033 and 0.046 for the power spectrum, and 0.035 and 0.042 for the bispectrum, respectively. 
The differences remain small in all cases, and the increase from the training to the test set is modest, indicating stable generalization across the cosmological grid. 
These figures therefore serve as a supplementary consistency check and support the conclusion that the model captures not only the mean behavior of the summary statistics, but also their distribution-level structure. While the distribution-level comparison of summary statistics provides a cosmologically motivated validation of the generated maps, direct pixel-level metrics for assessing the full conditional distribution, such as MIRA \cite{sharief2026mira}, offer a complementary perspective and will be an interesting direction for future work.
\section{Conclusion} \label{sec:conclusion}
We present a residual label-conditional flow matching framework for cosmological mass map generation conditioned on cosmological parameters. 
We demonstrate our method on 2D mass maps conditioned on $\Omega_m$ and $\sigma_8$, following previous work \cite{perraudin2021emulation}.
Our model substantially improves statistical fidelity of the generated maps and accelerates training convergence.

We evaluate the model using a comprehensive suite of quantitative metrics. 
For low-order statistics such as pixel histograms, peak histograms, and the power spectrum, the agreement is excellent, with typical discrepancies below 0.01 or $1\%$. 
For higher-order statistics, including Minkowski functionals and the bispectrum, our generated maps also achieve strong agreement, generally below $5\%$. 
The Structural Similarity Index (SSIM) and correlation matrix also show good agreement: the overall SSIM significance is well below unity, and the Frobenius-norm contrast of the correlation matrices is typically below $3\%$ for most cosmologies. 
We observe that errors are smallest near the center of the parameter grid, and that the degradation at the boundaries is now greatly reduced, with edge points exhibiting only a mild deterioration compared to the central region, which is expected because there is naturally less surrounding support in label space near the boundaries.

Overall, our results demonstrate that conditional flow matching, when combined with a residual-space formulation and a well-designed data-driven prior, provides a powerful and efficient framework for generating cosmological mass maps with high-fidelity non-Gaussian structure. 
These design choices substantially reduce optimization difficulties while improving generalization across cosmologies. 
They also lead to faster training and more stable performance at both boundary and interpolated parameter points, thereby enhancing the practical usability of the model. 
Compared to the GAN baseline of \cite{perraudin2021emulation}, our model achieves consistently smaller fractional differences across all evaluated summary statistics, while requiring substantially fewer training epochs to reach stable validation performance.

Looking ahead, several directions appear promising for extending this work. 
Jointly generating mass maps and shear fields could enable more realistic forward modeling of weak-lensing surveys. 
Incorporating redshift-dependent conditioning could enable tomographic mass map generation and reconstruction \cite{yiu2022tomographic,boruah2024gansky, boruah2024bayesian, boruah2025diffusion}.
Using full-sky simulation suites such as CosmoGridV1 \citep{kacprzak2023cosmogridv1} would allow us to train on spherical maps and to include a larger set of cosmological parameters, following recent spherical mass map emulators \citep{yiu2022tomographic,boruah2024gansky}.
Finally, integrating simulation-based inference techniques or uncertainty quantification could allow the model to serve directly as a likelihood-free tool for cosmological parameter estimation \citep{alsing2018massive,jeffrey2025dark,zeghal2025simulation}. 
We leave these directions for future work.

\acknowledgments
We acknowledge excellent previous work by Nathanaël Perraudin on the GAN emulator and the well-documented codes/data from that project, which is at the highest standard of reproducibility.

\bibliographystyle{JHEP}
\bibliography{biblio}

@article{bartelmann2001weak,
  title={Weak gravitational lensing},
  author={Bartelmann, Matthias and Schneider, Peter},
  journal={Physics Reports},
  volume={340},
  number={4-5},
  pages={291--472},
  year={2001},
  publisher={Elsevier}
}

@article{kilbinger2015cosmology,
  title={Cosmology with cosmic shear observations: a review},
  author={Kilbinger, Martin},
  journal={Reports on Progress in Physics},
  volume={78},
  number={8},
  pages={086901},
  year={2015},
  publisher={IOP Publishing}
}

@article{de2013kilo,
  title={{The Kilo-Degree Survey}},
  author={de Jong, Jelte TA and Verdoes Kleijn, Gijs A and Kuijken, Konrad H and Valentijn, Edwin A and KiDS and Astro-WISE Consortiums},
  journal={Experimental Astronomy},
  volume={35},
  number={1},
  pages={25--44},
  year={2013},
  publisher={Springer}
}

@article{dark2016dark,
  title={{The Dark Energy Survey}: more than dark energy--an overview},
  author={Dark Energy Survey Collaboration: and Abbott, T and Abdalla, FB and Aleksi{\'c}, J and Allam, S and Amara, A and Bacon, D and Balbinot, E and Banerji, M and Bechtol, K and others},
  journal={Monthly Notices of the Royal Astronomical Society},
  volume={460},
  number={2},
  pages={1270--1299},
  year={2016},
  publisher={Oxford University Press}
}

@article{aihara2018first,
  title={First data release of the {Hyper Suprime-Cam Subaru Strategic Program}},
  author={Aihara, Hiroaki and Armstrong, Robert and Bickerton, Steven and Bosch, James and Coupon, Jean and Furusawa, Hisanori and Hayashi, Yusuke and Ikeda, Hiroyuki and Kamata, Yukiko and Karoji, Hiroshi and others},
  journal={Publications of the Astronomical Society of Japan},
  volume={70},
  number={SP1},
  pages={S8},
  year={2018},
  publisher={Oxford University Press}
}

@article{laureijs2011euclid,
  title={Euclid definition study report},
  author={Laureijs, Rene and Amiaux, J{\'e}r{\^o}me and Arduini, S and Augueres, J-L and Brinchmann, J and Cole, R and Cropper, M and Dabin, C and Duvet, L and Ealet, AJAPA and others},
  journal={arXiv preprint arXiv:1110.3193},
  year={2011}
}

@article{ivezic2019lsst,
  title={{LSST}: from science drivers to reference design and anticipated data products},
  author={Ivezi{\'c}, {\v{Z}}eljko and Kahn, Steven M and Tyson, J Anthony and Abel, Bob and Acosta, Emily and Allsman, Robyn and Alonso, David and AlSayyad, Yusra and Anderson, Scott F and Andrew, John and others},
  journal={The Astrophysical Journal},
  volume={873},
  number={2},
  pages={111},
  year={2019},
  publisher={IOP Publishing}
}

@article{jain2000ray,
  title={Ray-tracing Simulations of Weak Lensing by Large-Scale Structure},
  author={Jain, Bhuvnesh and Seljak, Uro{\v{s}} and White, Simon},
  journal={The Astrophysical Journal},
  volume={530},
  number={2},
  pages={547},
  year={2000},
  publisher={IOP Publishing}
}

@article{takahashi2017full,
  title={Full-sky gravitational lensing simulation for large-area galaxy surveys and cosmic microwave background experiments},
  author={Takahashi, Ryuichi and Hamana, Takashi and Shirasaki, Masato and Namikawa, Toshiya and Nishimichi, Takahiro and Osato, Ken and Shiroyama, Kosei},
  journal={The Astrophysical Journal},
  volume={850},
  number={1},
  pages={24},
  year={2017},
  publisher={IOP Publishing}
}

@article{petri2016sample,
  title={Sample variance in weak lensing: How many simulations are required?},
  author={Petri, Andrea and Haiman, Zolt{\'a}n and May, Morgan},
  journal={Physical Review D},
  volume={93},
  number={6},
  pages={063524},
  year={2016},
  publisher={APS}
}

@article{harnois2019cosmic,
  title={Cosmic shear covariance matrix in {$w$CDM}: cosmology matters},
  author={Harnois-Deraps, Joachim and Giblin, Benjamin and Joachimi, Benjamin},
  journal={Astronomy \& Astrophysics},
  volume={631},
  pages={A160},
  year={2019},
  publisher={EDP Sciences}
}

@article{ribli2019weak,
  title={Weak lensing cosmology with convolutional neural networks on noisy data},
  author={Ribli, Dezs{\H{o}} and Pataki, B{\'a}lint {\'A}rmin and Zorrilla Matilla, Jos{\'e} Manuel and Hsu, Daniel and Haiman, Zolt{\'a}n and Csabai, Istv{\'a}n},
  journal={Monthly Notices of the Royal Astronomical Society},
  volume={490},
  number={2},
  pages={1843--1860},
  year={2019},
  publisher={Oxford University Press}
}

@article{fluri2022full,
  title={Full {$w$CDM} analysis of {KiDS-1000} weak lensing maps using deep learning},
  author={Fluri, Janis and Kacprzak, Tomasz and Lucchi, Aurelien and Schneider, Aurel and Refregier, Alexandre and Hofmann, Thomas},
  journal={Physical Review D},
  volume={105},
  number={8},
  pages={083518},
  year={2022},
  publisher={APS}
}

@article{sharma2024comparative,
  title={A comparative study of cosmological constraints from weak lensing using convolutional neural networks},
  author={Sharma, Divij and Dai, Biwei and Seljak, Uro{\v{s}}},
  journal={Journal of Cosmology and Astroparticle Physics},
  volume={2024},
  number={08},
  pages={010},
  year={2024},
  publisher={IOP Publishing}
}

@article{mustafa2019cosmogan,
  title={{CosmoGAN}: creating high-fidelity weak lensing convergence maps using generative adversarial networks},
  author={Mustafa, Mustafa and Bard, Deborah and Bhimji, Wahid and Luki{\'c}, Zarija and Al-Rfou, Rami and Kratochvil, Jan M},
  journal={Computational Astrophysics and Cosmology},
  volume={6},
  number={1},
  pages={1},
  year={2019},
  publisher={Springer}
}

@article{perraudin2021emulation,
  title={Emulation of cosmological mass maps with conditional generative adversarial networks},
  author={Perraudin, Nathana{\"e}l and Marcon, Sandro and Lucchi, Aurelien and Kacprzak, Tomasz},
  journal={Frontiers in Artificial Intelligence},
  volume={4},
  pages={673062},
  year={2021},
  publisher={Frontiers Media SA}
}

@article{yiu2022tomographic,
  title={A tomographic spherical mass map emulator of the {KiDS-1000} survey using conditional generative adversarial networks},
  author={Yiu, Timothy Wing Hei and Fluri, Janis and Kacprzak, Tomasz},
  journal={Journal of Cosmology and Astroparticle Physics},
  volume={2022},
  number={12},
  pages={013},
  year={2022},
  publisher={IOP Publishing}
}

@article{remy2023probabilistic,
  title={Probabilistic mass-mapping with neural score estimation},
  author={Remy, Benjamin and Lanusse, Francois and Jeffrey, Niall and Liu, Jia and Starck, J-L and Osato, Ken and Schrabback, Tim},
  journal={Astronomy \& Astrophysics},
  volume={672},
  pages={A51},
  year={2023},
  publisher={EDP Sciences}
}

@article{aoyama2025denoising,
  title={Denoising weak lensing mass maps with diffusion model: systematic comparison with generative adversarial network},
  author={Aoyama, Shohei D. and Osato, Ken and Shirasaki, Masato},
  journal={arXiv preprint arXiv:2505.00345},
  year={2025}
}

@article{boruah2025diffusion,
  title={Diffusion-based mass map reconstruction from weak lensing data},
  author={Boruah, Supranta S and Jacob, Michael and Jain, Bhuvnesh},
  journal={Physical Review D},
  volume={111},
  number={8},
  pages={083542},
  year={2025},
  publisher={APS}
}

@article{dinh2016density,
  title={Density estimation using {Real NVP}},
  author={Dinh, Laurent and Sohl-Dickstein, Jascha and Bengio, Samy},
  journal={arXiv preprint arXiv:1605.08803},
  year={2016}
}

@article{kingma2018glow,
  title={{Glow}: generative flow with invertible {1x1} convolutions},
  author={Kingma, Durk P and Dhariwal, Prafulla},
  journal={Advances in neural information processing systems},
  volume={31},
  year={2018}
}

@article{papamakarios2021normalizing,
  title={Normalizing flows for probabilistic modeling and inference},
  author={Papamakarios, George and Nalisnick, Eric and Rezende, Danilo Jimenez and Mohamed, Shakir and Lakshminarayanan, Balaji},
  journal={Journal of Machine Learning Research},
  volume={22},
  number={57},
  pages={1--64},
  year={2021}
}

@article{lipman2022flow,
  title={Flow matching for generative modeling},
  author={Lipman, Yaron and Chen, Ricky TQ and Ben-Hamu, Heli and Nickel, Maximilian and Le, Matt},
  journal={arXiv preprint arXiv:2210.02747},
  year={2022}
}

@article{liu2022flow,
  title={Flow straight and fast: Learning to generate and transfer data with rectified flow},
  author={Liu, Xingchao and Gong, Chengyue and Liu, Qiang},
  journal={arXiv preprint arXiv:2209.03003},
  year={2022}
}

@article{tong2023improving,
  title={Improving and generalizing flow-based generative models with minibatch optimal transport},
  author={Tong, Alexander and Fatras, Kilian and Malkin, Nikolay and Huguet, Guillaume and Zhang, Yanlei and Rector-Brooks, Jarrid and Wolf, Guy and Bengio, Yoshua},
  journal={arXiv preprint arXiv:2302.00482},
  year={2023}
}

@article{diao2025detecting,
  title={Detecting modeling bias with continuous time flow models on weak lensing maps},
  author={Diao, Kangning and Dai, Biwei and Seljak, Uro{\v{s}}},
  journal={Journal of Cosmology and Astroparticle Physics},
  volume={2025},
  number={08},
  pages={004},
  year={2025},
  publisher={IOP Publishing}
}

@article{zeghal2025bridging,
  title={Bridging simulators with conditional optimal transport},
  author={Zeghal, Justine and Remy, Benjamin and Hezaveh, Yashar and Lanusse, Francois and Levasseur, Laurence Perreault},
  journal={arXiv preprint arXiv:2510.24631},
  year={2025}
}

@article{kannan2025cosmoflow,
  title={{CosmoFlow}: scale-aware representation learning for cosmology with flow matching},
  author={Kannan, Sidharth and Qiu, Tian and Cuesta-Lazaro, Carolina and Jeong, Haewon},
  journal={arXiv preprint arXiv:2507.11842},
  year={2025}
}

@article{kollovieh2024flow,
  title={Flow matching with gaussian process priors for probabilistic time series forecasting},
  author={Kollovieh, Marcel and Lienen, Marten and L{\"u}dke, David and Schwinn, Leo and G{\"u}nnemann, Stephan},
  journal={arXiv preprint arXiv:2410.03024},
  year={2024}
}

@article{issachar2025designing,
  title={Designing a Conditional Prior Distribution for Flow-Based Generative Models},
  author={Issachar, Noam and Salama, Mohammad and Fattal, Raanan and Benaim, Sagie},
  journal={arXiv preprint arXiv:2502.09611},
  year={2025}
}

@article{wu2025flowdesign,
  title={{FlowDesign}: Improved design of antibody {CDRs} through flow matching and better prior distributions},
  author={Wu, Jun and Kong, Xiangzhe and Sun, Ningguan and Wei, Jing and Shan, Sisi and Feng, Fuli and Wu, Feng and Peng, Jian and Zhang, Linqi and Liu, Yang and others},
  journal={Cell Systems},
  volume={16},
  number={6},
  pages={101270},
  year={2025},
  publisher={Elsevier}
}

@article{tamosiunas2021investigating,
  title={Investigating cosmological GAN emulators using latent space interpolation},
  author={Tamosiunas, Andrius and Winther, Hans A and Koyama, Kazuya and Bacon, David J and Nichol, Robert C and Mawdsley, Ben},
  journal={Monthly Notices of the Royal Astronomical Society},
  volume={506},
  number={2},
  pages={3049--3067},
  year={2021},
  publisher={Oxford University Press}
}

@article{chen2018neural,
  title={Neural ordinary differential equations},
  author={Chen, Ricky TQ and Rubanova, Yulia and Bettencourt, Jesse and Duvenaud, David K},
  journal={Advances in neural information processing systems},
  volume={31},
  year={2018}
}

@article{fluri2019cosmological,
  title={Cosmological constraints with deep learning from {KiDS-450} weak lensing maps},
  author={Fluri, Janis and Kacprzak, Tomasz and Lucchi, Aurelien and Refregier, Alexandre and Amara, Adam and Hofmann, Thomas and Schneider, Aurel},
  journal={arXiv preprint arXiv:1906.03156},
  year={2019}
}

@article{taruya2002lognormal,
  title={Lognormal property of weak-lensing fields},
  author={Taruya, Atsushi and Takada, Masahiro and Hamana, Takashi and Kayo, Issha and Futamase, Toshifumi},
  journal={The Astrophysical Journal},
  volume={571},
  number={2},
  pages={638},
  year={2002},
  publisher={IOP Publishing}
}

@article{clerkin2017testing,
  title={Testing the lognormality of the galaxy and weak lensing convergence distributions from {Dark Energy Survey} maps},
  author={Clerkin, Lucinda and Kirk, Donnacha and Manera, M and Lahav, O and Abdalla, F and Amara, Adam and Bacon, D and Chang, C and Gaztanaga, E and Hawken, A and others},
  journal={Monthly Notices of the Royal Astronomical Society},
  volume={466},
  number={2},
  pages={1444--1461},
  year={2017},
  publisher={Oxford University Press}
}

@article{boruah2022map,
  title={Map-based cosmology inference with lognormal cosmic shear maps},
  author={Boruah, Supranta S and Rozo, Eduardo and Fiedorowicz, Pier},
  journal={Monthly Notices of the Royal Astronomical Society},
  volume={516},
  number={3},
  pages={4111--4122},
  year={2022},
  publisher={Oxford University Press}
}

@article{wang2004image,
  title={Image quality assessment: from error visibility to structural similarity},
  author={Wang, Zhou and Bovik, Alan C and Sheikh, Hamid R and Simoncelli, Eero P},
  journal={IEEE Transactions on Image Processing},
  volume={13},
  number={4},
  pages={600--612},
  year={2004},
  publisher={IEEE}
}

@article{petri2016mocking,
  title={Mocking the weak lensing universe: the {LensTools} {Python} computing package},
  author={Petri, Andrea},
  journal={Astronomy and Computing},
  volume={17},
  pages={73--79},
  year={2016},
  publisher={Elsevier}
}

@article{kacprzak2023cosmogridv1,
  title={{CosmoGridV1}: a simulated {$\Lambda$CDM} theory prediction for map-level cosmological inference},
  author={Kacprzak, Tomasz and Fluri, Janis and Schneider, Aurel and Refregier, Alexandre and Stadel, Joachim},
  journal={Journal of Cosmology and Astroparticle Physics},
  volume={2023},
  number={02},
  pages={050},
  year={2023},
  publisher={IOP Publishing}
}

@article{sharief2026mira,
  title={{MIRA}: A Score for Conditional Distribution Accuracy and Model Comparison},
  author={Sharief, Sammy and Zeghal, Justine and Barco, Gabriel Missael and Lemos, Pablo and Hezaveh, Yashar and Perreault-Levasseur, Laurence},
  journal={arXiv preprint arXiv:2605.02014},
  year={2026}
}

@article{boruah2024gansky,
  title={{GANSky}: fast curved-sky weak lensing simulations using generative adversarial networks},
  author={Boruah, Supranta S and Fiedorowicz, Pier and Garcia, Rafael and Coulton, William R and Rozo, Eduardo and Fabbian, Giulio},
  journal={arXiv preprint arXiv:2406.05867},
  year={2024}
}

@article{boruah2024bayesian,
  title={Bayesian mass mapping with weak lensing data using {KaRMMa}: validation with simulations and application to {Dark Energy Survey Year 3} data},
  author={Boruah, Supranta S and Fiedorowicz, Pier and Rozo, Eduardo},
  journal={Physical Review D},
  volume={110},
  number={2},
  pages={023524},
  year={2024},
  publisher={APS}
}

@article{alsing2018massive,
  title={Massive optimal data compression and density estimation for scalable, likelihood-free inference in cosmology},
  author={Alsing, Justin and Wandelt, Benjamin and Feeney, Stephen},
  journal={Monthly Notices of the Royal Astronomical Society},
  volume={477},
  number={3},
  pages={2874--2885},
  year={2018},
  publisher={Oxford University Press}
}

@article{jeffrey2025dark,
  title={{Dark Energy Survey Year 3} results: likelihood-free, simulation-based {$w$CDM} inference with neural compression of weak-lensing map statistics},
  author={Jeffrey, Niall and Whiteway, Lorne and Gatti, Marco and Williamson, J and Alsing, Justin and Porredon, Anna and Prat, Judit and Doux, Cyrille and Jain, Bhuvnesh and Chang, Chihway and others},
  journal={Monthly Notices of the Royal Astronomical Society},
  volume={536},
  number={2},
  pages={1303--1322},
  year={2025},
  publisher={Oxford University Press}
}

@article{zeghal2025simulation,
  title={Simulation-based inference benchmark for weak lensing cosmology},
  author={Zeghal, Justine and Lanzieri, Denise and Lanusse, Fran{\c{c}}ois and Boucaud, Alexandre and Louppe, Gilles and Aubourg, Eric and Bayer, Adrian E},
  journal={Astronomy \& Astrophysics},
  volume={699},
  pages={A327},
  year={2025},
  publisher={EDP Sciences}
}

@inproceedings{wang2003multiscale,
  title={Multiscale structural similarity for image quality assessment},
  author={Wang, Zhou and Simoncelli, Eero P and Bovik, Alan C},
  booktitle={The Thirty-Seventh Asilomar Conference on Signals, Systems \& Computers, 2003},
  volume={2},
  pages={1398--1402},
  year={2003},
  organization={IEEE}
}

\appendix
\section{Network architecture}\label{app:network}

\paragraph{Dataset.}
The dataset used in this work is described in Section~\ref{sec:dataset}. Each mass map has a pixel dimension of $1\times 128\times 128$.
We select 46 cosmologies (552,000 samples) as the training set and 
11 cosmologies (132,000 samples) as the test set. 
For the original maps, we first apply the lognormal transformation to the 
training data, then we standardize the maps using the global mean and 
standard deviation computed over all training pixels, as described in 
Section~\ref{sec:dataset}.

\paragraph{Predictor for conditional mean and variance.}
After data preprocessing, we designed a predictor to estimate the conditional mean $\mu_0(y)$ and standard deviation $\sigma_0(y)$ of the prior distribution, as described in Section~\ref{sec:rlcfm}. Because the mass maps are approximately statistically homogeneous and isotropic at a fixed cosmology, the one-point statistics are invariant under translations and rotations, implying that the ensemble mean is (to a good approximation) independent of pixel position. Consistent with
this expectation, we find empirically that pixel-wise averaging over
many realizations at the same cosmology yields an almost spatially
constant mean map. We therefore model only its global mean, reducing the task to predicting a single scalar $\mu_0(y)$, which is then broadcast to a constant $1\times128\times128$ mean map. Thus, the main task of our predictor is to predict two scalar values: $\mu_0(y)$ and $\sigma_0(y)$. This prediction model is trained only on the 46 training cosmologies, and the values for held-out test cosmologies are obtained by interpolation without using any test maps.

We model $\sigma_0(y)$ with a two-stage scheme.  First, a low-order Polynomial Ridge regressor is fit to $\{(y_i,\sigma_i)\}$ to obtain a smooth baseline $\sigma_{\rm base}(y)$. Second, we fit a Gradient Boosted Decision Trees (GBDT) regressor to the residuals and add a bounded correction
\begin{align}
\sigma_0(y) \;=\; \sigma_{base}(y) \;+\;
\Delta\sigma(y)\,,
\end{align}
where the correction is clipped as
$|\Delta\sigma|\le \kappa_\sigma \max\{\sigma_{\rm base}(y),\epsilon_\sigma\}$
with $\kappa_\sigma=0.02$ to ensure positivity and numerical stability.

We model $\mu_0(y)$ analogously. As a baseline, we train a Gaussian Process Regressor (GPR) with a squared-exponential kernel and additive white noise on $\{(y_i,\mu_i)\}$ (after standardizing inputs and targets), yielding $\mu_{\rm base}(y)$. A second-stage GBDT predicts residuals, giving
\begin{align}
\mu_0(y) \;=\; \mu_{\rm base}(y) \;+\; \Delta\mu(y),
\label{mu_pre}
\end{align}
where $|\Delta\mu|\le \kappa_\mu \max\{\sigma_0(y),\epsilon_\mu\}$ with $\kappa_\mu=0.02$ to prevent unphysical excursions. During sampling, we broadcast $\mu_0(y)$ to a constant mean map and use $\sigma_0(y)$ to set the noise schedule in the flow matching process.

For hyperparameters, we set $\alpha=1.2$ in Eq.~(\ref{sigma0}) and $\beta=0.04$ in Eq.~(\ref{sigma1}). The choice of $\alpha>1$ increases the coverage of the initial distribution, which improves learning of large-scale structure.
\begin{table}[t]
\centering
\small
\begin{tabular}{lcc}
\toprule
\textbf{Layer} & \textbf{Input shape} & \textbf{Output shape} \\
\midrule
\multicolumn{3}{l}{\textbf{Downsampling layers}} \\
\quad Conv2d & [128, 1, 128, 128] & [128, 32, 128, 128] \\
\quad Cond ResBlock & [128, 32, 128, 128] & [128, 32, 128, 128] \\
\quad Cond ResBlock & [128, 32, 128, 128] & [128, 32, 128, 128] \\
\quad Downsample & [128, 32, 128, 128] & [128, 32, 64, 64] \\
\quad Cond ResBlock & [128, 32, 64, 64] & [128, 64, 64, 64] \\
\quad Cond ResBlock & [128, 64, 64, 64] & [128, 64, 64, 64] \\
\quad Downsample & [128, 64, 64, 64] & [128, 64, 32, 32] \\
\quad Cond ResBlock & [128, 64, 32, 32] & [128, 128, 32, 32] \\
\quad Cond ResBlock & [128, 128, 32, 32] & [128, 128, 32, 32] \\
\quad Downsample & [128, 128, 32, 32] & [128, 128, 16, 16] \\
\quad Cond ResBlock & [128, 128, 16, 16] & [128, 128, 16, 16] \\
\quad Cond AttentionBlock & [128, 128, 16, 16] & [128, 128, 16, 16] \\
\quad Cond ResBlock & [128, 128, 16, 16] & [128, 128, 16, 16] \\
\quad Cond AttentionBlock & [128, 128, 16, 16] & [128, 128, 16, 16] \\
\midrule
\multicolumn{3}{l}{\textbf{Middle layers}} \\
\quad Cond ResBlock & [128, 128, 16, 16] & [128, 128, 16, 16] \\
\quad Cond AttentionBlock & [128, 128, 16, 16] & [128, 128, 16, 16] \\
\quad Cond ResBlock & [128, 128, 16, 16] & [128, 128, 16, 16] \\
\midrule
\multicolumn{3}{l}{\textbf{Upsampling layers}} \\
\quad Cond ResBlock & [128, 256, 16, 16] & [128, 128, 16, 16] \\
\quad Cond AttentionBlock & [128, 128, 16, 16] & [128, 128, 16, 16] \\
\quad Cond ResBlock & [128, 256, 16, 16] & [128, 128, 16, 16] \\
\quad Cond AttentionBlock & [128, 128, 16, 16] & [128, 128, 16, 16] \\
\quad Cond ResBlock & [128, 256, 16, 16] & [128, 128, 16, 16] \\
\quad Cond AttentionBlock & [128, 128, 16, 16] & [128, 128, 16, 16] \\
\quad Upsample & [128, 128, 16, 16] & [128, 128, 32, 32] \\
\quad Cond ResBlock & [128, 256, 32, 32] & [128, 128, 32, 32] \\
\quad Cond ResBlock & [128, 256, 32, 32] & [128, 128, 32, 32] \\
\quad Cond ResBlock & [128, 192, 32, 32] & [128, 128, 32, 32] \\
\quad Upsample & [128, 128, 32, 32] & [128, 128, 64, 64] \\
\quad Cond ResBlock & [128, 192, 64, 64] & [128, 64, 64, 64] \\
\quad Cond ResBlock & [128, 128, 64, 64] & [128, 64, 64, 64] \\
\quad Cond ResBlock & [128, 96, 64, 64] & [128, 64, 64, 64] \\
\quad Upsample & [128, 64, 64, 64] & [128, 64, 128, 128] \\
\quad Cond ResBlock & [128, 96, 128, 128] & [128, 32, 128, 128] \\
\quad Cond ResBlock & [128, 64, 128, 128] & [128, 32, 128, 128] \\
\quad Cond ResBlock & [128, 64, 128, 128] & [128, 32, 128, 128] \\
\quad Conv2d & [128, 32, 128, 128] & [128, 1, 128, 128] \\
\bottomrule
\end{tabular}
\caption{Summary of the conditional U-Net architecture used as the velocity network in the RLCFM model. The network operates on $128\times128$ single channel maps and is composed of four downsampling stages, a middle bottleneck, and a symmetric four stage upsampling path with skip connections. Time and label conditioning are implemented through dedicated embeddings that are injected into the conditional residual and attention blocks. The table lists the main layers and their input and output tensor shapes.}
\label{tab:unet}
\end{table}

\paragraph{U-Net model.}
We use a U-Net architecture for the velocity network, operating on 128 × 128 single channel maps, described in Table \ref{tab:unet}. In particular,  we designed a time and label embedding block, and specifically for label embedding, we incorporated physical features. First, we apply a Fourier feature mapping to the parameters with a learnable scale $\gamma$ that controls the feature bandwidth. To prevent $\gamma$ from becoming excessively large and inducing high frequency oscillations, we apply a small weight decay to this parameter.
Beyond the Fourier representation, we further augment the input with physics-inspired features, including an $S_8$ term with learnable gating, logarithmic terms, and physically motivated ratios, thereby embedding prior structural knowledge of the underlying problem into the representation. The time and label embedding block is then embedded separately into the resblock and attention block, i.e., the CondResBlock and CondAttentionBlock in  Table \ref{tab:unet}.

The network comprises four
downsampling stages with channel widths $32\!\rightarrow\!64\!\rightarrow\!128\!\rightarrow\!128$,
followed by a bottleneck at $16\times16$, and a symmetric upsampling path
with skip connections from the corresponding downsampling stages.
Each downsampling stage consists of two conditioning ResNet blocks and a
downsampling operation; at the lowest resolution ($16\times16$) we
additionally apply conditioning attention blocks to capture long range
dependencies. The middle (bottleneck) consists of a conditioning ResNet
block, a conditioning Attention block, and a second conditioning ResNet
block. The upsampling path mirrors the encoder: at each resolution we
concatenate the skip features with the current activations, apply
conditioning ResNet blocks (with conditioning attention blocks at
$16\times16$), and then upsample back to the original resolution. A final
$3\times3$ convolution maps the $32$-channel output back to a single
channel.

\paragraph{Training.}
We train the model with the AdamW optimizer, using a starting learning
rate of $2\times10^{-4}$ that is decayed with a cosine schedule to
$3\times10^{-5}$ by the end of training. We use a batch size of 128,
with each batch constructed by mixing samples from multiple
training cosmologies. An exponential moving average (EMA) of the network
weights is maintained throughout training, and we use the EMA weights
for all subsequent sampling.

\paragraph{Sampling.}
We generate samples using the \texttt{dopri5} ODE solver.  Since the
dynamics are integrated in the residual space $r$, we recover the field
in the original space using the mean prediction $\mu_0(y)$ from
Eq.~(\ref{mu_pre}), i.e.
\begin{align}
\tilde x \;=\; r + \mu_0(y).
\end{align}
Finally, we apply the inverse of the data preprocessing transformations. First, we invert the global standardization in Eq.~(\ref{g}),
\begin{align}
&x = \sigma_g\tilde{x} + \mu_g.
\end{align}
Since the stored maps correspond to the shifted convergence field, the inverse logarithmic transform gives
\begin{align}
&\kappa_{\rm shifted} = e^x.
\end{align}
All generated maps and summary statistics in this work are evaluated using this shifted field, consistent with the stored dataset. On average, it takes about 4 minutes to generate 5000 maps on an NVIDIA A100 GPU.
\section{Experiments and analysis} \label{app:ablation}

\subsection{Comparison with GAN}
We compare our mass map emulator against the GAN-based emulator of \cite{perraudin2021emulation}, using the same dataset and data split, and the same validation metrics. Overall, our RLCFM model achieves consistently smaller fractional differences than the GAN across all summary statistics. We mainly show the corresponding comparisons for the pixel value distribution, power spectrum, the correlation matrix, and the bispectrum.

For the pixel value distribution, the GAN shows normalized Wasserstein-1 distances at the level of a few percent, with the largest values approaching or exceeding $5\%$. In contrast, the RLCFM model remains below $1\%$ for 
 nearly all training and test points.
For the power spectrum, the GAN typically shows fractional differences of order a few percent, reaching or exceeding $5\%$ at several boundary parameter points. In contrast, the RLCFM model yields fractional differences below $1\%$ for most test points, with only a small subset reaching $\sim 2\%$.
For the correlation matrix, the GAN often exhibits approximate differences at the \(\sim 5\%\)-\(15\%\) level, with the largest deviations reaching $\sim 40\%$. The RLCFM model performs substantially better: aside from a few points reaching $\sim 9\%$, most points cluster around $\sim 3\%$.
For the bispectrum, the GAN typically reaches fractional differences of order $\sim 10\%$, while several edge points reach or exceed $20\%$. By comparison, the RLCFM model achieves differences below $5\%$ for most test points, with only a few points falling in the range $5\%$-$10\%$.

In addition to improved accuracy, our RLCFM model is more training-efficient. While the GAN baseline in \cite{perraudin2021emulation} is trained for 40 epochs, RLCFM already reaches stable validation metrics after approximately $\sim 15$ epochs, with no noticeable improvement thereafter. This reduces the required training budget and accelerates model development.

Overall, these results demonstrate that our RLCFM emulator provides a clear improvement over the GAN baseline. Across all four summary statistics, it not only reduces the typical fractional differences, but also suppresses the worst case errors at boundary cosmologies, where the GAN degrades most strongly. This indicates that our approach better preserves both two-point and higher-order statistics, yielding more reliable mass map generation over the full \((\Omega_m,\sigma_8)\) grid. Moreover, RLCFM reaches stable validation performance within a substantially smaller epoch budget, further improving computational efficiency.

\subsection{Ablation study}

\begin{figure}
  \centering
  \includegraphics[width=0.85\linewidth]{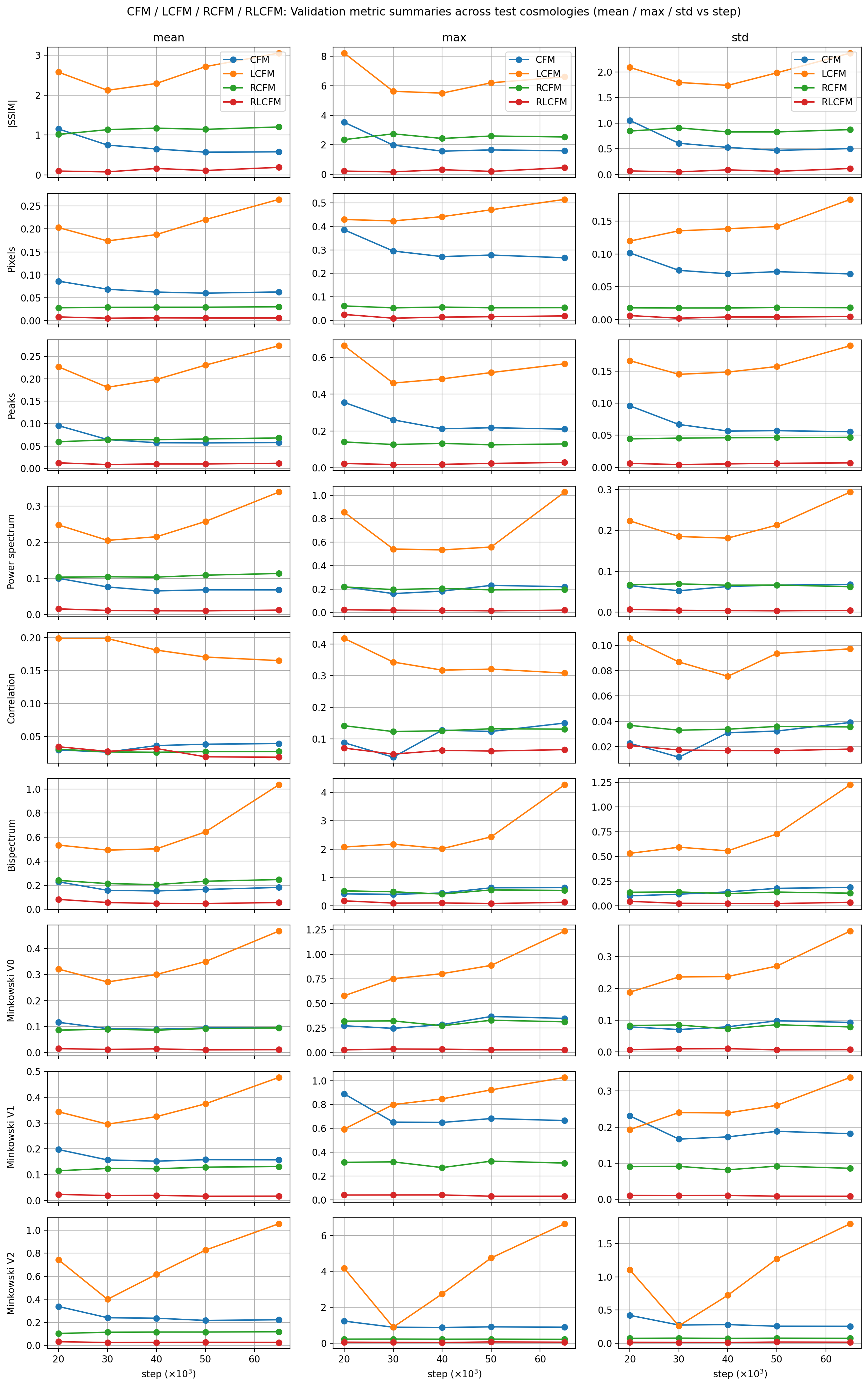}
  \caption{Validation metric summaries across 11  test cosmologies (lower is better). Columns report mean, max (worst case), and std across the 11 test cosmological parameter points as a function of training steps. Colors denote CFM (blue), LCFM (orange), RCFM (green), and RLCFM (red). RLCFM achieves the lowest mean/max/std across metrics, while LCFM consistently performs worst, highlighting the benefit of combining residual-space training with the label-specific prior for robust interpolation.}
  \label{summary_mean_max_std}
\end{figure}

To assess the impact of the label-specific prior and residual-space reparameterization on interpolation generalization across cosmological parameters, we conduct ablation experiments on 11  test cosmological parameter points (lower is better), and report the results in Fig.~\ref{summary_mean_max_std}.

\paragraph{Baseline (CFM).}
We use the standard CFM, where all labels share a standard Gaussian prior and the model is trained in the original data space.

\paragraph{Ablation A (LCFM; label-specific prior only).}
We replace the shared prior with a label-specific prior, while still training the model in the original data space.

\paragraph{Ablation B (RCFM; residual space only).}
We keep the shared prior as in the baseline, but train the model in the residual space via a residual-space reparameterization (mean subtraction) $r = x - \mu_0(y)$.

\paragraph{Ablation C (RLCFM; full).}
We adopt the label-specific prior and train the model in the residual space using the same reparameterization $r = x - \mu_0(y)$.

Figure~\ref{summary_mean_max_std} summarizes interpolation-focused performance on 11  test cosmological parameter points. We report the validation metrics (as in Section~\ref{sec:metrics}) aggregated across these test cosmologies, including the mean, maximum (worst case), and standard deviation (cross cosmology variability) as a function of training steps, where lower values indicate better performance. Across all statistics, including $|\mathrm{SSIM}|$, pixel, peak, power spectrum, correlation, bispectrum, and Minkowski functionals, LCFM (label-specific prior only) consistently exhibits the largest mean values, the worst max values, and the highest variability, indicating degraded interpolation generalization and reduced robustness at test interpolation points. A plausible explanation is that the neural vector field shares parameters across all labels, whereas the label-specific prior induces different source distributions and thus different starting locations in the state space for different $y$. As a result, the model has to approximate a family of label-conditional vector fields whose dependence on $y$ may become less smooth, especially when the source distributions for neighboring labels have limited overlap. This can increase the effective functional complexity of $v_\theta(t,\cdot,y)$ in the label direction and lead to poor interpolation generalization between the training labels, which is reflected by degraded performance at test interpolation points.

\begin{figure}
  \centering
  \includegraphics[width=0.8\linewidth]{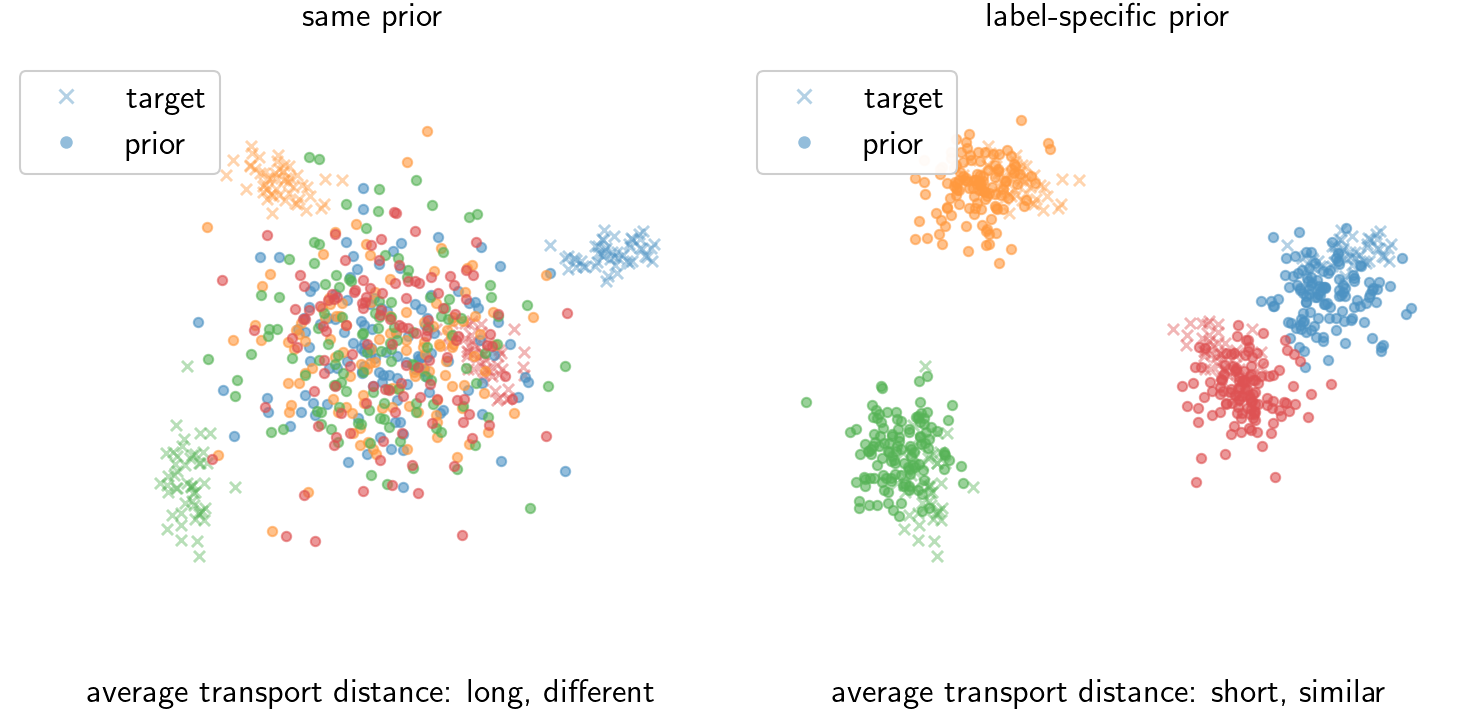} 
  \caption{\textbf{Motivation for a label-specific source prior.} 
Left: using a shared source prior \(p(x_0)\) for all labels, source samples (dots; ``prior'') can be far from the corresponding label-conditional targets \(q(x| y)\) (crosses; ``target''), leading to long and different transport. 
Right: a label-specific prior \(p(x_0|y)\) places source samples closer to their corresponding targets on average, yielding shorter and more similar transport across labels.}
  \label{label_prior}
\end{figure}

In contrast, residual-space reparameterization alone (RCFM) removes the per-label mean component via $r=x-\mu_0(y)$. This aligns the conditional distributions around $r=0$ and reduces label-dependent variability in the input space, thereby making the family of label-conditional vector fields more consistent across $y$ under a shared parameter network. As a result, the conditional learning problem becomes easier and the model can devote its capacity to modeling the residual, non-Gaussian structures that are more transferable across nearby cosmologies. Correspondingly, RCFM improves most non-Gaussian statistics, for example peak counts and Minkowski functionals, relative to the CFM baseline.

Importantly, combining the label-specific prior with residual-space training (RLCFM) gives the best overall performance. It achieves the lowest mean and worst case values and the smallest standard deviations across all reported validation metrics, and it remains stable over training steps. This indicates that the two components are complementary. Residual-space training first removes the main label-dependent mean shifts, making the learning problem more similar across labels and improving interpolation with shared network parameters. The label-specific prior then brings the starting distribution closer to the residual target distribution for each $y$ by using a physically motivated variance (see Fig.~\ref{label_prior}). This provides a better scaled initialization for each cosmology and effectively injects data-informed prior knowledge into the model, improving training stability and leading to stronger generalization at interpolation points. Together, these two choices lead to more accurate and more robust interpolation across the cosmological parameter space, which motivates the joint design.

\section{Theorem proofs and reparameterization} \label{app:proof}

\subsection{Theorems and proofs} \label{app:theorem}

\textbf{Theorem 1.} \textit{Given the label-conditional vector field $u_t(x|x_1, y)$ that generates $p_t(x|x_1,y)$, the marginal conditional vector field $u_t(x| y)$ in Eq.~\eqref{eq:u_t(x|y)} generates the
marginal conditional probability path in Eq.~\eqref{eq:p_t(x|y)}.}
\medskip
\begin{proof}
We assume sufficient smoothness and integrability to interchange differentiation with respect to \(t\), as well as the spatial divergence with respect to \(x\), with integration over \(x_1\). We also assume that \(p_t(x|y)>0\) on the region of interest. Because $u_t(x|x_1,y)$ generates $p_t(x|x_1,y)$, they satisfy the continuity equation Eq.~\eqref{eq:continuity},
\begin{equation}
\frac{\partial p_t(x|x_1,y)}{\partial t} = -\nabla_x\cdot(p_t(x|x_1,y)u_t(x|x_1,y)).
\end{equation}
Then we can prove that $u_t(x|y)$ and $p_t(x|y)$ also satisfy the continuity equation
\begin{align}
\frac{\partial p_t(x|y)}{\partial t} &= \frac{\partial}{\partial t}\int p_t(x|x_1,y)q(x_1|y)dx_1 \nonumber \\
&= \int \frac{\partial}{\partial t}p_t(x|x_1,y) \cdot q(x_1|y)dx_1 \nonumber \\
&= -\int \nabla_x\cdot(p_t(x|x_1,y)u_t(x|x_1,y)) \cdot q(x_1|y)dx_1 \nonumber \\
&= -\nabla_x\cdot \int p_t(x|x_1,y)u_t(x|x_1,y)q(x_1|y)dx_1 \nonumber \\
&= -\nabla_x\cdot(p_t(x|y)u_t(x|y)),
\end{align}
where in the first equation we used Eq.~\eqref{eq:p_t(x|y)}, and in the last equation we used Eq.~\eqref{eq:u_t(x|y)}. Therefore $u_t(x|y)$ we defined generates $p_t(x|y)$. 
\end{proof}

\bigskip
\noindent\textbf{Theorem 2.} \textit{
Assume that \(p_t(x\mid y)>0\) for all \(x\in\mathbb{R}^d\),
\(y\in\mathcal{Y}\), and \(t\in[0,1]\). Then the label-marginal
flow matching objective
\[
\mathcal{L}_{\mathrm{FM}}^{y}(\theta)
=
\mathbb{E}_{t\sim\mathcal{U}[0,1],\,q(y),\,p_t(x\mid y)}
\bigl\|v_\theta(t,x,y)-u_t(x\mid y)\bigr\|^2,
\]
and the label-conditional flow matching objective
\[
\mathcal{L}_{\mathrm{LCFM}}(\theta)
=
\mathbb{E}_{t\sim\mathcal{U}[0,1],\,q(x_1,y),\,p_t(x\mid x_1,y)}
\bigl\|v_\theta(t,x,y)-u_t(x\mid x_1,y)\bigr\|^2,
\]
differ only by a constant independent of \(\theta\). Hence,
\[
\nabla_\theta \mathcal{L}_{\mathrm{FM}}^{y}(\theta)
=
\nabla_\theta \mathcal{L}_{\mathrm{LCFM}}(\theta).
\]
}
\medskip
\begin{proof}
For notational simplicity, we omit the explicit dependence on \(t\) in
some intermediate expressions. Using the bilinearity of the squared norm,
we have
\[
\begin{aligned}
\bigl\|v_\theta(t,x,y)-u_t(x|y)\bigr\|^2
=
\bigl\|v_\theta(t,x,y)\bigr\|^2
-2\bigl\langle v_\theta(t,x,y),u_t(x| y)\bigr\rangle  
+\bigl\|u_t(x|y)\bigr\|^2 ,
\end{aligned}
\]
and
\[
\begin{aligned}
\bigl\|v_\theta(t,x,y)-u_t(x| x_1,y)\bigr\|^2
=
\bigl\|v_\theta(t,x,y)\bigr\|^2
-2\bigl\langle v_\theta(t,x,y),u_t(x| x_1,y)\bigr\rangle 
+\bigl\|u_t(x|x_1,y)\bigr\|^2 .
\end{aligned}
\]
The last terms in the two expressions are independent of \(\theta\).
It remains to show that the two \(\theta\)-dependent terms are identical
under the corresponding expectations.

First, using the definition of the label-marginal probability path in Eq.~(\ref{eq:p_t(x|y)}),
\[
p_t(x| y)
=
\int p_t(x| x_1,y) q(x_1| y)\,dx_1,
\]
we obtain
\[
\begin{aligned}
\mathbb{E}_{q(y),\,p_t(x| y)}
\bigl\|v_\theta(t,x,y)\bigr\|^2 
&=
\int q(y)p_t(x| y)
\bigl\|v_\theta(t,x,y)\bigr\|^2\,dx\,dy  \\
&=
\int q(y)
\left[
\int p_t(x| x_1,y)q(x_1| y)\,dx_1
\right]
\bigl\|v_\theta(t,x,y)\bigr\|^2\,dx\,dy  \\
&=
\int q(y)q(x_1| y)p_t(x| x_1,y)
\bigl\|v_\theta(t,x,y)\bigr\|^2\,dx\,dx_1\,dy  \\
&=
\int q(x_1,y)p_t(x| x_1,y)
\bigl\|v_\theta(t,x,y)\bigr\|^2\,dx\,dx_1\,dy  \\
&=
\mathbb{E}_{q(x_1,y),\,p_t(x| x_1,y)}
\bigl\|v_\theta(t,x,y)\bigr\|^2. 
\end{aligned}
\]
Second, using the definition of the label-marginal vector field in Eq.~(\ref{eq:u_t(x|y)}),
\[
u_t(x| y)
=
\int u_t(x|x_1,y)
\frac{p_t(x|x_1,y)q(x_1|y)}
     {p_t(x| y)}
\,dx_1 ,
\]
we have
\[
\begin{aligned}
\mathbb{E}_{q(y),\,p_t(x|y)}
\bigl\langle v_\theta(t,x,y),u_t(x|y)\bigr\rangle  
&=
\int q(y)p_t(x|y)
\left\langle
v_\theta(t,x,y),
\int u_t(x|x_1,y)
\frac{p_t(x|x_1,y)q(x_1|y)}
     {p_t(x|y)}
\,dx_1
\right\rangle
dx\,dy  \\
&=
\int q(y)
\left\langle
v_\theta(t,x,y),
\int u_t(x| x_1,y)
p_t(x| x_1,y)q(x_1|y)
\,dx_1
\right\rangle
dx\,dy  \\
&=
\int q(y)q(x_1| y)p_t(x| x_1,y)
\bigl\langle
v_\theta(t,x,y),u_t(x| x_1,y)
\bigr\rangle
\,dx\,dx_1\,dy  \\
&=
\int q(x_1,y)p_t(x| x_1,y)
\bigl\langle
v_\theta(t,x,y),u_t(x| x_1,y)
\bigr\rangle
\,dx\,dx_1\,dy  \\
&=
\mathbb{E}_{q(x_1,y),\,p_t(x| x_1,y)}
\bigl\langle
v_\theta(t,x,y),u_t(x| x_1,y)
\bigr\rangle.
\end{aligned}
\]
Therefore, the \(\|v_\theta(t,x,y)\|^2\) terms and the cross terms are
identical in \(\mathcal{L}_{\mathrm{FM}}^{y}(\theta)\) and
\(\mathcal{L}_{\mathrm{LCFM}}(\theta)\). The only remaining terms,
\[
\mathbb{E}_{q(y),\,p_t(x|y)}
\bigl\|u_t(x| y)\bigr\|^2
\]
and
\[
\mathbb{E}_{q(x_1,y),\,p_t(x| x_1,y)}
\bigl\|u_t(x| x_1,y)\bigr\|^2 
\]
do not depend on \(\theta\). Hence, the two objectives differ only by a
constant independent of \(\theta\). Consequently,
\[
\nabla_\theta \mathcal{L}_{\mathrm{FM}}^{y}(\theta)
=
\nabla_\theta \mathcal{L}_{\mathrm{LCFM}}(\theta).
\]
\end{proof}

\subsection{Reparameterization of the Gaussian probability path}\label{app:rep}

In Section \ref{2.3}, we define a label-specific Gaussian probability path $p_t(x|x_1,y)$. To evaluate the Label-Conditional Flow Matching (LCFM) loss efficiently, we reparameterize the path. Given an initial position $x_0$, the associated flow $\psi_t(x_0|x_1,y)$ that transports $x_0$ to $x_t$ is given by
\begin{equation}
x_t|_{x_1,y} = \psi_t(x_0|x_1,y) = \mu_t + \sigma_t \left( \frac{x_0 - \mu_0}{\sigma_0} \right).
\end{equation}
By applying the reparameterization trick and setting $z_0 = (x_0 - \mu_0)/\sigma_0$, we obtain $z_0 \sim \mathcal{N}(0, I)$. The flow can then be simplified as
\begin{equation}
x_t|_{x_1,y} = \psi_t(z_0|x_1,y) = \mu_t + \sigma_t z_0.
\end{equation}
Taking the time derivative along this reparameterized path gives the
pathwise velocity
\begin{equation}
\dot{x}_t = \dot\mu_t + \dot\sigma_t z_0,
\end{equation}
where $\dot\mu_t$ and $\dot\sigma_t$ denote the time derivatives of the mean and standard deviation, respectively. Substituting this parameterization into the LCFM objective allows us to train the neural network to regress the target vector field by sampling from a standard normal distribution, leading to the final objective function in Eq.~\eqref{loss_LCFM}.




\end{document}